\documentclass[12pt,aps,prd,eqsecnum]{revtex4}
\usepackage{amssymb,amsmath,amsthm,graphicx,amscd}
\usepackage{enumerate,color,comment,ulem}
\usepackage[justification=Justified,format=plain]{caption}
\usepackage{bm,floatrow,wrapfig,subcaption}
\usepackage{xcolor}

\usepackage[hidelinks]{hyperref}

\usepackage{orcidlink}

\usepackage{xcolor}



\begin{document}

\title{Atom-Field-Medium Interactions III: Quantum Field-mediated Entanglement between Two Atoms near a Conducting Surface}
\author{Jen-Tsung Hsiang\orcidlink{0000-0002-9801-208X}}
\email{cosmology@gmail.com}
\affiliation{College of Electrical Engineering and Computer Science, National Taiwan University of Science and Technology, Taipei City, Taiwan 106, R.O.C.}
\author{Bei-Lok Hu\orcidlink{0000-0003-2489-9914}}\email{blhu@umd.edu}
\affiliation{Maryland Center for Fundamental Physics and Joint Quantum Institute,  University of Maryland, College Park, Maryland 20742, USA}
\date{July 25, 2025}

\newpage
\begin{abstract}
This third paper in this series continues the investigation of  atom-field interactions in the presence of a conductor or a dielectric medium, focusing on quantum information related basic issues such as decoherence and entanglement.  Here we consider the entanglement between two atoms with internal degrees of freedom modeled by a harmonic oscillator, with varying separations between them and varying distances between them and a conducting surface. These are configurations familiar in the Casimir-Polder effect, but the behavior of atom-surface entanglement is quite different from the well-studied behavior of field-induced forces. For one, while  the attractive force between an atom and a conducting surface increases as they come  closer, the entanglement between the atom and the quantum field actually decreases as the atom gets closer to the conductor, as shown in \cite{Rong,AFD2}.  We show how different factors play out, ranging from the coupling between the atoms and the field to the coupling between the atoms, going beyond the weak coupling restrictions often found necessary in the literature. Gathering our results for the entanglement dependence on each variable concerned, we can provide a spatial topography of quantum  entanglement, thus enabling a visualized understanding of the behavior of quantum field-mediated entanglement. In particular we can quantify the definition of a three-dimensional \textit{entanglement domain} between the two atoms,  how it varies with their coupling, their separation and their distances from the conducting surface, and for practical applications, how to exercise effective control of the entanglement between two atoms by changing these parameters. Our findings are expected to be useful for studies of atom-field-medium interactions in vacuum and surface physics.
\end{abstract}

\maketitle
\tableofcontents
\section{Introduction}
The present work continues to explore certain foundational theoretical issues of atom-field-medium interactions, focusing on quantum field-mediated entanglement between two atoms near a boundary. It adheres to the spirit of this series, namely, starting from first principles, using microphysics models, employing methodology respecting self-consistency, exploring strong coupling effects, and seeking exact solutions.  In Paper I \cite{AFD1}  we derived the graded influence action for  a system of $N$ neutral atoms whose internal degrees of freedom, modeled by harmonic oscillators (which provides the dipole moment),  interact with a quantum scalar field in the presence of a linear dielectric medium. In Paper II \cite{AFD2}  we derived the quantum Langevin equation for the nonMarkovian dynamics of this system interacting with the dielectric-modified quantum field, and presented the analytic forms of the covariance matrix elements of this system when it is in a steady state. As an illustration of their applications we analyzed the quantum entanglement between one such atom and a dielectric half space in terms of the purity function.  

Quantum field-mediated effects are of fundamental theoretical significance and practical relevance. Let us parse the three ingredients in the subtitle of this paper: quantum field theoretical (QFT) effects, boundary effects and quantum-field mediated entanglement between atoms, and describe their broader relevance in succession.  A prime example of  how boundaries and topology \cite{QFTop,QFTop1,QFTop2,QFTop3,QFTop4} enter in quantum field theory  is the Casimir effect \cite{Casimir,Casimir1,Casimir2,Casimir3,Casimir4}.  Dynamics and curvature are the major themes  in the well-established field of QFT in curved spacetime (CST) \cite{QFCS,QFCS1,QFCS2},  with important examples of dynamical Casimir effect \cite{DCE} and cosmological particle creation \cite{CPC}. The effects due to an event horizon are manifested in the Unruh effect \cite{Unr76} as thermal radiance experienced by an accelerated observer, and the Hawking effect \cite{Haw74} of thermal radiation emanated from a black hole. Finally, concerning entanglement, not only is it a uniquely quantum feature \cite{Schroedinger} resting in the bedrock of quantum mechanics,  it is also the primary resource in quantum information processing which powers the second quantum revolution in motion. Adding quantum informational issues to the consideration of quantum field theory, one would enter the realm of relativistic quantum information \cite{RQI}, which deals with fundamental issues from special relativity \cite{Peres} to quantum gravity \cite{EntQG}. In the vase expense of current frontier research subjects, quantum entanglement plays a distinctly important role.  

While many of these effects of fundamental significance belong to the hardly accessible realms of black holes and the early universe,  analog gravity \cite{AnalogG} experiments bring them closer to the laboratory. Among them, those based on atoms, mirrors and optics (AMO) occupy an important place.  The Unruh-DeWitt detector \cite{Unr76,Unr761,DeW79} in the Unruh effect could be a two-level atom  or a harmonic atom. Another famous analog of the Hawking effect is the moving mirror \cite{MovMir,MovMir1,MovMir2,MovMir3,MovMir4}.  Studies of entanglement between two atoms mediated by a quantum field including vacuum fluctuations \cite{2AEntT,2AEntT1,2AEntT2,2AEntT3,2AEntT4} have both practical experimental significance \cite{2AEntE,2AEntE1,2AEntE2}  as well as implications for cosmology and black hole physics via analog gravity.  In the AMO context this is related to exploring the quantum information issues of the Casimir-Polder effect \cite{CasPol}.  For the practical aspects of atom-atom entanglement near a dielectric medium in actual experimental settings we refer to the recent paper of \cite{Kanu24} whose Introduction explains the motivations and provides the background with ample references. 

In this paper we consider the entanglement between two harmonic atoms  anywhere above  the conducting surface. These are configurations familiar in the Casimir-Polder (CP) effect, but the behavior of atom-surface entanglement is quite different from the well-studied behavior of field-induced forces. For one, while  the attractive force between an atom and a conducting surface increases as they come  closer, the entanglement between the atom and the quantum field actually decreases as the atom gets closer to the conductor, as shown in \cite{Rong,AFD2}.   While the effects related to vacuum energy arising from quantum field fluctuations such as Casimir, Casimir-Polder and dynamical Casimir and dynamical Casimir-Polder effects have been studied for a long time  and quite well understood, investigations of the quantum informational aspects of these effects are still in a developing stage.   

In this vein, the entanglement between two uniformly accelerated UDW detectors have been studied by some authors \cite{Ent2UD,Ent2UD1,Ent2UD2,Ent2UD3} \footnote{Here, we suggest exercising proper judgment in accepting results obtained from familiar yet inadequate theories such as the time-dependent perturbation theory,  which is only valid for weak coupling and for short times, or under oft-invoked yet over-simplified assumptions, such as the random wave or Born-Markov approximations~\cite{KapTjo23}. Beware that even the most popularly important master equations have limitations, e.g., the Caldeira-Leggett master equation \cite{CalLeg83} is not completely positive \cite{HomaCL}, and the Lindblad \cite{Lindblad}--GKS \cite{GKS} master equation, though completely positive and trace preserving, fails at very low temperatures~\cite{LinbPatho,LinbPatho1,LinbPatho2}.}.  At this stage of our investigation we treat only stationary atoms,  for the purpose of uncovering as much as possible the hitherto scantily explored non-perturbative effects, based on exact solutions, as exemplified in \cite{LinHu06,LinHu07},  and,  in the way how the atom/mirror-field entanglement was treated, in the style as exemplified by \cite{LinHu09,MOF2}, or the entanglement between two atoms when field-mediation competes with direct coupling, as in  \cite{HHPRD16}.  Only after these results are rigorously established and well understood will we feel assured enough to start to explore quantum informational issues in moving atoms and  moving mirrors.    

As a step forward from \cite{LinHu09,HHPRD16} we shall consider field-mediated entanglement between two atoms in the presence of a conducting surface. 
We show how different factors play out, ranging from the coupling between the atoms and the field to the coupling between the atoms, going beyond the weak coupling restrictions often found necessary in the literature. Gathering our results for the entanglement dependence on each variable concerned, we can provide a spatial topography of quantum  entanglement, thus enabling a visualized understanding of the behavior of quantum field-mediated entanglement. In particular we can quantify the definition of a three-dimensional \textit{entanglement domain} between the two atoms,  how it varies with their coupling, their separation and their distances from the conducting surface, and for practical applications, how to exercise effective control of the entanglement between two atoms by changing these parameters. Our findings are expected to be useful for studies of atom-field-medium interactions in vacuum and surface physics.

A brief description of our goal, the set up and the approach we employ:    
Our goal is to investigate the spatial dependence of the late-time entanglement between two interacting harmonic atoms, whose internal degrees of freedom are coupled to a common ambient massless scalar field in the presence of a perfectly conducting boundary. The ambient field is assumed to be initially in the vacuum state, and interacts simultaneously with the internal degrees of freedom of two atoms situated at distinct spatial positions. The evolutions of each atom's internal dynamics is determined by solving the coupled Heisenberg equations governing both atoms' internal degrees of freedom and the quantum field. From these solutions, the covariance matrix elements associated with the atomic degrees of freedom can be computed. These elements form the basis to construct measures that determine the degree of entanglement between the two atoms.

\subsection{Key Findings}

The key findings of our investigation can be grouped under under two categories:  
1) A quantifiable notion of ``entanglement domain (ED)";  
2) How the boundary (BE) affects the entanglement. 
We enumerate them in the following:

\subsubsection{Entanglement Domain} 

\begin{enumerate}
    \item For the present model, we can identify a well-defined spherical entanglement domain at late times. When both atoms are located within this domain, their internal degrees of freedom become entangled.
    \item The size of the domain is determined by the coupling strength, the specific form of atom-field interaction, and the initial state of the ambient field.
    \item In general, the shape of the entanglement domain gets compressed when the atoms are situated near the boundary.
\end{enumerate}

\subsubsection{Boundary Effects} 
\begin{enumerate}
    \item Keeping the position of atom 1 fixed, the entanglement between the two atoms typically improves when atom 2 is located closer to the boundary.
    \item However, subtleties emerge when atom 1 is positioned very close to the boundary. The generation of quantum entanglement in this regime reflects a competition between these two effects: the  field-mediated mutual interaction, which promotes coherence between the atoms, and quantum noise from field fluctuations, which induces decoherence. Both effects are suppressed near the boundary due to the Dirichlet condition requiring the field amplitude to vanish there. The net outcome depends on which effect dominates.  
    \item Numerical results show that quantum entanglement between the two atoms is weakened in the immediate vicinity of the perfectly conducting boundary. In this region,  sharp transitions appear in the entanglement behavior, indicating a shift in dominance between enhancement from mutual interaction and attenuation due to decoherence. When the atoms are very close to the boundary, entanglement between them could be subtle as the results of theoretical calculations are likely model-dependent. One should consider incorporating physical material properties to warrant a closer comparison  with experiments.
\end{enumerate}
Additional conceptual ambiguities arise from the renormalization procedure, when both atoms are placed extremely close to the idealized perfectly conducting boundary. This further complicates the physical interpretation of results in such configurations.

This paper is organized as follows. In Sec.~II, we examine the general influence of a perfectly conducting boundary on the internal dynamics of an atom through its coupling to the ambient quantum field. In Sec.~III, we identify the region of dynamical instability that arises from strong field-mediated interactions when the two atoms are in close proximity. Section~IV is devoted to analyzing the spatial structure of the quantum entanglement between the two atoms, referred to as the entanglement domain, under the current configuration. In Sec.~V, we investigate how this domain depends on factors such as coupling strength, spatial separation, and the presence of the boundary. We conclude with a summary of our main results.

\section{Boundary Effect on an Atom's Internal Dynamics}\label{S:verued}

Consider two neutral atoms in three-dimensional half-space ($z>0$), bounded by a perfectly conducting plate at $z=0$. The internal degrees of freedom (idf) of these two atoms, modeled by quantum harmonic oscillators, are minimally coupled to an ambient massless quantum scalar field $\phi$, satisfying the Dirichlet boundary condition $\phi=0$ on the boundary~\footnote{In the case where the atomic internal degrees of freedom are coupled to the scalar field via derivative coupling, analogous to the coupling between an electric charge and an electromagnetic field, this type of minimal coupling under consideration can alternatively be interpreted as arising from  applying order reduction to  {both} the dissipative and fluctuation backactions of the field  {in the open-system framework}~\cite{Rohrlich,Yaghjian,Erber61,MS74,FOC91,PJM82,GLR10,LS19,HL08,GHW09,BDR19,HH22,AFD2}.}. 

\begin{figure}
    \centering
    \includegraphics[width=0.5\linewidth]{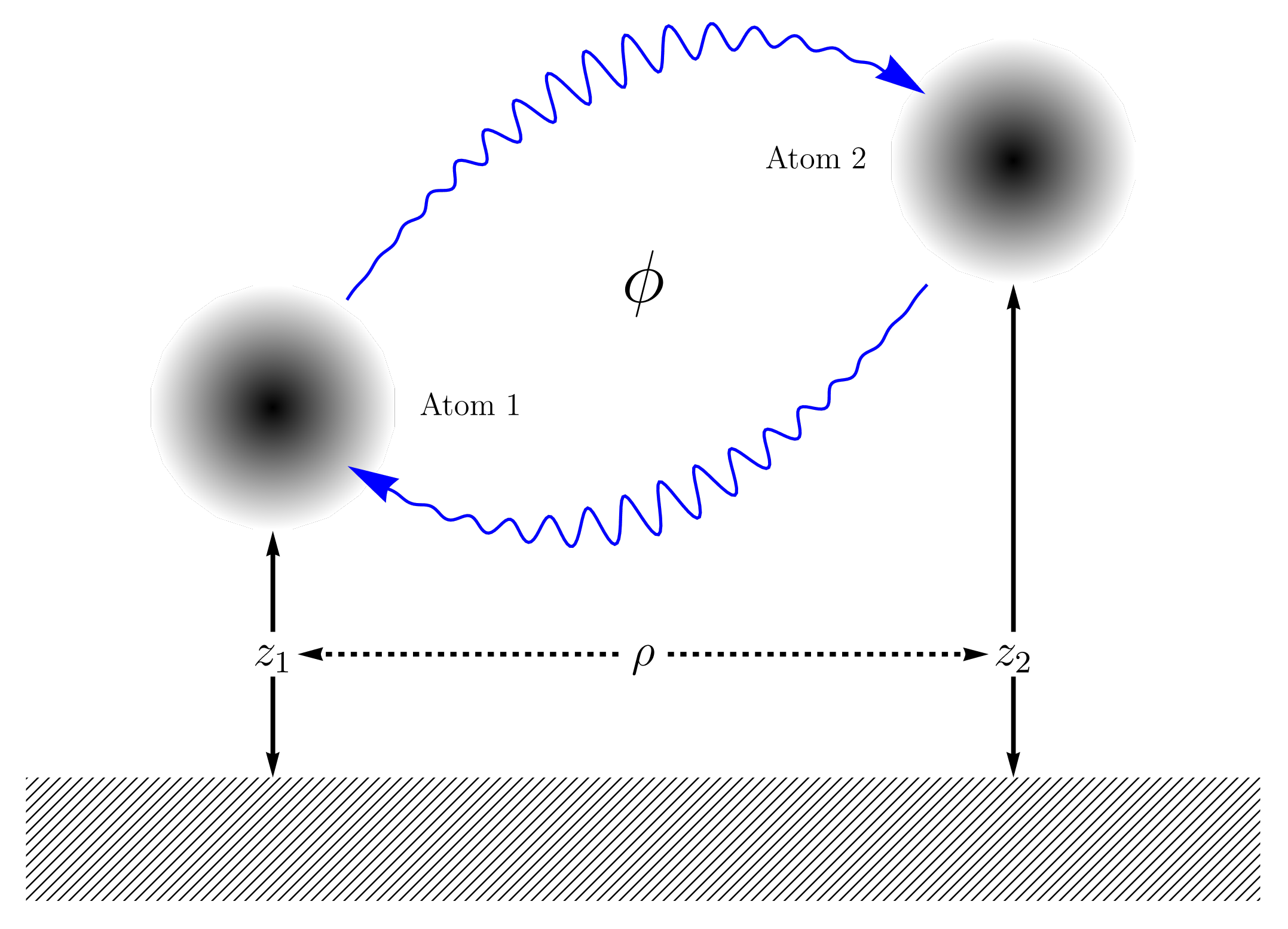}
    \caption{The spatial arrangement of two interacting atoms outside a perfect conductor, denoted by the hatched area. Both atoms interact with a common quantum field $\phi$, modified by the boundary condition depending on the surface properties of the conductor. The blue wavy curves denote the nonMarkovian  mutual influence  between the two atoms mediated by the field.}
    \label{Fi:afi}
\end{figure}

We assume that atom 1 is located at the position $\bm{x}_1=(\bm{x}_{1,\perp},z_1)$ and atom 2 at $\bm{x}_2=(\bm{x}_{2,\perp},z_2)$, with $\perp$ referring to the {transverse components normal} to the $z$ axis, and $z_1$, $z_2>0$. Their horizontal separation is denoted by $\rho=\lvert \bm{x}_{1,\perp}-\bm{x}_{2,\perp}\rvert$, as shown in Fig.~\ref{Fi:afi}. We wish to introduce the notion of entanglement domain and show how it can be quantified, and to study the effects of the boundary on the quantum entanglement between two atoms,

Let $\hat{\chi}_i$ denote the operator of the canonical variable of atom $i$. Solving a simultaneous set of Heisenberg equations governing the internal degrees of freedom of two atoms and the ambient scalar field, we obtain the reduced equation of motion for $\hat{\chi}_i$, 
\begin{align}
    \ddot{\hat{\chi}}_1(t)+\omega^2_{\textsc{b}}\,\hat{\chi}_1(t)-\frac{e^2}{m}\sum_{j=1}^2\int_0^t\!ds\;G^{(\phi_h)}_{\textsc{r}}(\bm{x}_1,t;\bm{x}_j,s)\,\hat{\chi}_j(s)&=\frac{e}{m}\,\hat{\phi}_h(\bm{x}_1,t)\,,\label{E:dkbfo1}\\
    \ddot{\hat{\chi}}_2(t)+\omega^2_{\textsc{b}}\,\hat{\chi}_2(t)-\frac{e^2}{m}\sum_{j=1}^2\int_0^t\!ds\;G^{(\phi_h)}_{\textsc{r}}(\bm{x}_2,t;\bm{x}_j,s)\,\hat{\chi}_j(s)&=\frac{e}{m}\,\hat{\phi}_h(\bm{x}_2,t)\,,\label{E:dkbfo2}
\end{align}
where we have assumed that both atoms are coupled to the field $\phi$ with the same coupling strength $e$, and that their internal degrees of freedom (dof) have identical natural frequency $\omega_{\textsc{b}}$ and mass $m$. The component $\hat{\phi}_h$ denotes the homogeneous solution of the wave equation that the full field $\hat{\phi}$ obeys,  often referred to as the `free' field. The two-point function $G^{(\phi_h)}_{\textsc{r}}$ is the retarded Green's function constructed by this free field (see Appendix A for a summary description.) Note that it is crucial to distinguish between the free field $\hat{\phi}_h$ and the full field $\hat{\phi}$; the latter includes the radiation field emitted by the atoms from the dynamics of their internal dof.

The dynamics described by Eqs.~\eqref{E:dkbfo1} and \eqref{E:dkbfo2} has been discussed with great details in Refs.~\cite{HHPRD16,HHPRD18,HAH22,AHH23}. Here,  subtleties arise due to the presence of the boundary. The two-point function $G^{(\phi_h)}_{\textsc{r}}(x,x')$ can be decomposed as a coherent superposition of the contributions of each atom and its image in unbounded space,  {located in position $\tilde{\bm{x}}'$}, as outlined in Appendix~A. Thus, the nonlocal term in either equation of motion appears deceptively simple. It actually contains four distinct contributions. Take Eq.~\eqref{E:dkbfo1} for example. The function $G^{(\phi_h)}_{\textsc{r}}(\bm{x}_1,t;\bm{x}_1,s)$ can be written as the sum of $G^{(\phi_h)}_{\textsc{r},0}(\bm{x}_1,t;\bm{x}_1,s)$ and $-G^{(\phi_h)}_{\textsc{r},0}(\bm{x}_1,t;\tilde{\bm{x}}_1,s)$. The former gives rise to the shift of oscillating frequency and the emergence of the damping term, as a consequence of radiation damping. The latter on the other hand is the delayed influence of atom 1 onto itself, mediated by the field reflected off the boundary, or equivalently interpreted as the induced nonMarkovian effect arising from the field, originated from the image of atom 1 at $\tilde{\bm{x}}_1$. The minus sign results from the Dirichlet boundary condition. Moreover, the kernel $G^{(\phi_h)}_{\textsc{r}}(\bm{x}_1,t;\bm{x}_2,t')$  accounts for the nonlocal influence from atom 2 and its image  {at $\tilde{\bm{x}}_2$ from an earlier instant $t'$}.

Since the Dirichlet boundary condition requires that the field vanishes on the boundary, the two-point function $G^{(\phi_h)}_{\textsc{r}}(x,x')$ is diminishingly small when one of the two points $x$, $x'$ lies in the immediate neighborhood of the boundary. This means that the internal degree of freedom of the atom is almost but not entirely decoupled from the field. As a result, it behaves nearly like a free harmonic oscillator with a slightly shifted oscillation frequency and negligible damping. Thus, the internal dynamics takes significantly longer relaxation times to reach equilibrium, compared to the situation where the atom is placed farther away from the boundary.

However, this observation does not readily imply that we can extrapolate the above observation to the case when the atom is right next to the conducting plate, and claim that the internal degree of freedom acts as a free oscillator there. Some subtle issues remain open in this case. First,   when the atom is far away from the plate, regularization and renormalization are usually necessary to obtain finite values of the parameters that describe the atoms' internal dynamics that interacts with an infinite number of field modes by Eqs.~\eqref{E:dkbfo1} and \eqref{E:dkbfo2}. But in contrast, the Dirichlet boundary implies the absence of the nonlocal influence in Eqs.~\eqref{E:dkbfo1} and \eqref{E:dkbfo2}, so when the atom is posited next to the plate, renormalization seems unnecessary, and we {might} directly work with the un-renormalized parameters of the atomic system. This prompts a revisit of the interpretation of the renormalization procedure. In addition, introduction of the cutoff scale implies that the Dirichlet boundary condition  {should be} implemented in the sense of effective-field theory, where the condition is enforced on field modes whose frequencies are lower than the cutoff frequency. The conducting plate is {treated as} transparent to the higher frequency modes.

\begin{figure}
    \centering
    \includegraphics[width=0.5\linewidth]{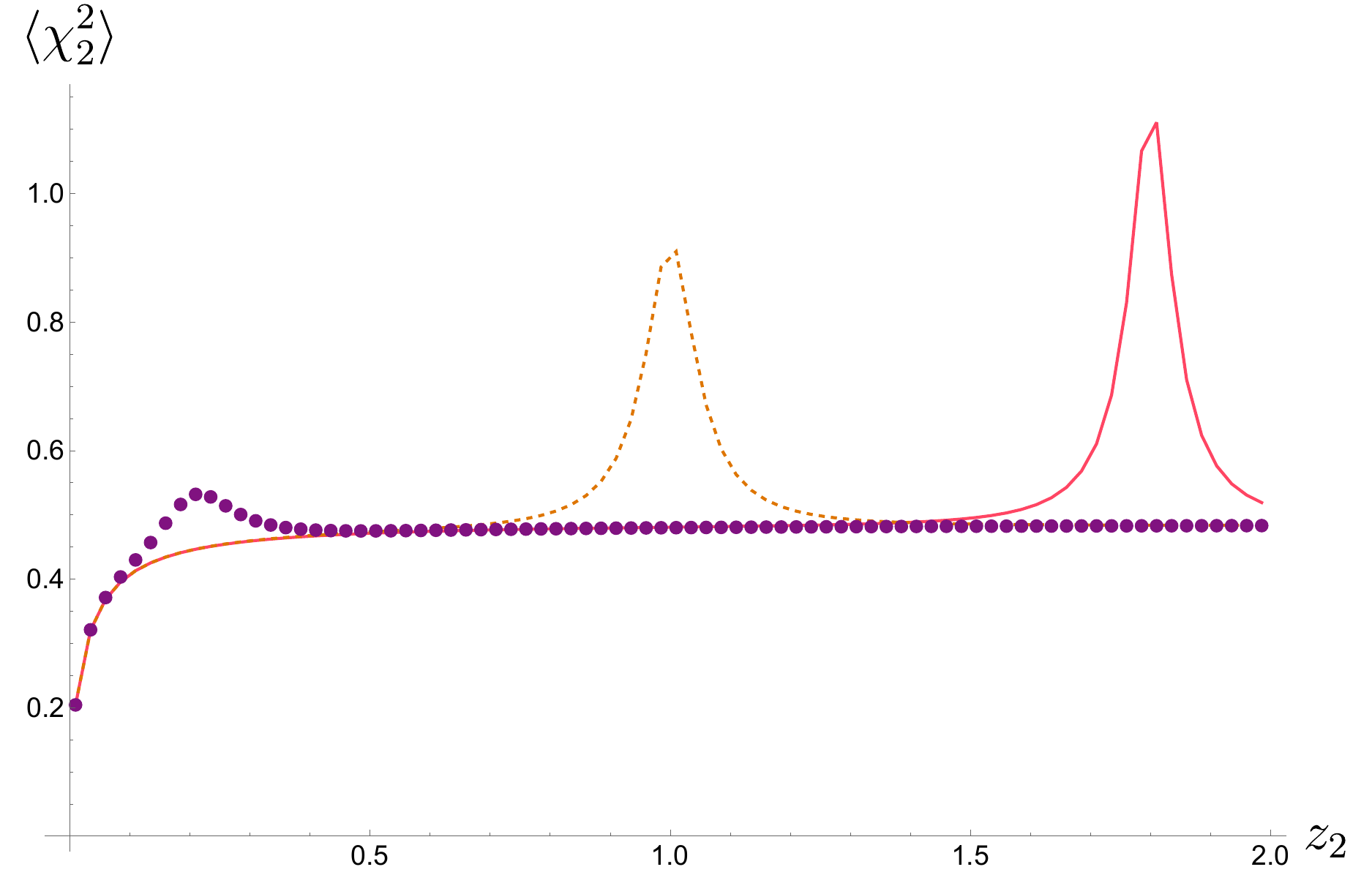}
    \caption{A typical example of the behavior of the expectation value of the operator of the canonical variable at late times for atom 2.  {The quantity $\langle\chi_2^2(\infty)\rangle$ measures uncertainty in the displacement of the internal degrees of freedom.} The horizontal axis $z_2$ denotes the vertical distance of atom 2 to the conducting plate. The curves exhibit peaks in the vicinity of atom 1, so the three curves correspond to the cases that  {the atom 1 is fixed at} $z_1=0.2$ (solid curve), $1$  {(thin dotted curve)} and $1.8$  {(thick dotted curve)} respectively (in units of wavelength associated with the renormalized atomic transition frequency). The horizontal separation between two atoms in this plot is $\rho=0.1$. Observe that the curves all move downward as $z$ approach the boundary. This is the consequence that in the vicinity of the plate the effects of the ambient field is suppressed by the boundary condition. This has a profound implication in the conceptual issues of the open-system framework.}
    \label{Fi:x2x2}
\end{figure}

The boundary effects on atomic internal dynamics is illustrated in Fig.~\ref{Fi:x2x2}, which shows results for three configurations where atom 1 is placed at the vertical distance $z_1=0.2$, $1$, and $1.8$ to the plate, in the units of wavelength associated with the renormalized atomic transition frequency, i.e., shifted oscillation frequency due to the atom-field interaction in the absence of the boundary. The horizontal axis $z_2$ denotes the vertical distance of atom 2 to the conducting plate. In this case, we choose the horizontal separation $\rho$ to be 0.1 in the same units. The vertical axis of the plot gives the uncertainty of the internal degree of freedom of atom 2 in its final equilibrium state. The peaks of the curves result from strong interaction when both atoms are sufficiently close; otherwise, these three curves more or less coincide.

Two notable features are observed in this plot. The height of the peaks decreases as atom 1  {is placed closer to} the boundary, indicating a weakening of interactions. Secondly, in the limit $z_2\to0$, the quantity $\langle\chi_2^2(\infty)\rangle$ drops to smaller values. This can be understood as follows. In the limit of weak coupling strength, the displacement uncertainty $\langle\chi_2^2(\infty)\rangle$ is inversely proportional to the square root of the shifted oscillation frequency $\omega_{r}$. Since the atom-field interaction under this type of coupling tends to produce a negative correction, the shifted oscillation frequency will be smaller than the original natural frequency.  Consequently, the oscillation frequency gradually increases as $z_2$ approaches the boundary.

\section{Stability Conditions for Internal Dynamics}

We aim to analyze the late-time entanglement between two neutral atoms due to the interaction of their internal degrees of freedom mediated by the ambient field in the presence of a flat, perfectly conducting boundary.  {Two relevant points here: First, what would be a useful and computable entanglement measure for the system under consideration? Negativity, von Neumann entropy and purity are computable and quantifiable measures of entanglement commonly invoked for bipartite Gaussian systems.  A summary of their properties is given in Appendix B.     Second,  the late-time behavior of these entanglement measures can be determined once the relevant information about the elements of the covariance matrix is available. But, one should ask, under what conditions in the dynamics of the internal degrees of freedom of the atoms interacting with a medium-modified quantum field would a simple enough late time behavior show up in  our system which is amenable to qualitative analysis~\cite{HHPRD18}?   We address this issue in this section. }

In the case of a single atom, its nonequilibrium internal dynamics is pretty well understood, and it is known to relax to an equilibrium state over a time scale determined by the inverse of the damping constant if the ambient field is initially prepared in a stationary state.  When multiple atoms are present, dynamical equilibration is no longer ensured. Describing the nonequilibrium evolution and establishing whether a final equilibrium state exists get more complicated.   

To obtain the time evolution for a system described by a set of simultaneous equations of motion like Eqs.~\eqref{E:dkbfo1} and \eqref{E:dkbfo2},  one can in principle convert them into algebraic equations using Laplace transformations,  assuming that the Laplace transforms exist. The solutions to the transformed equations are then transformed back to the time domain, yielding the full time evolution of $\chi_1$ and $\chi_2$. Fortunately, it is often unnecessary to solve the transformed equations explicitly. Instead, it suffices to identify the locations of the poles of these algebraic equations.  For clarity and convenience, this analysis is sometimes carried out in the frequency domain rather than in terms of the Laplace parameter, because the two are related via a Wick rotation. 

\begin{figure}
    \centering
    \hfill
    \begin{subfigure}[b]{0.46\textwidth}
         \centering
         \includegraphics[width=0.95\linewidth]{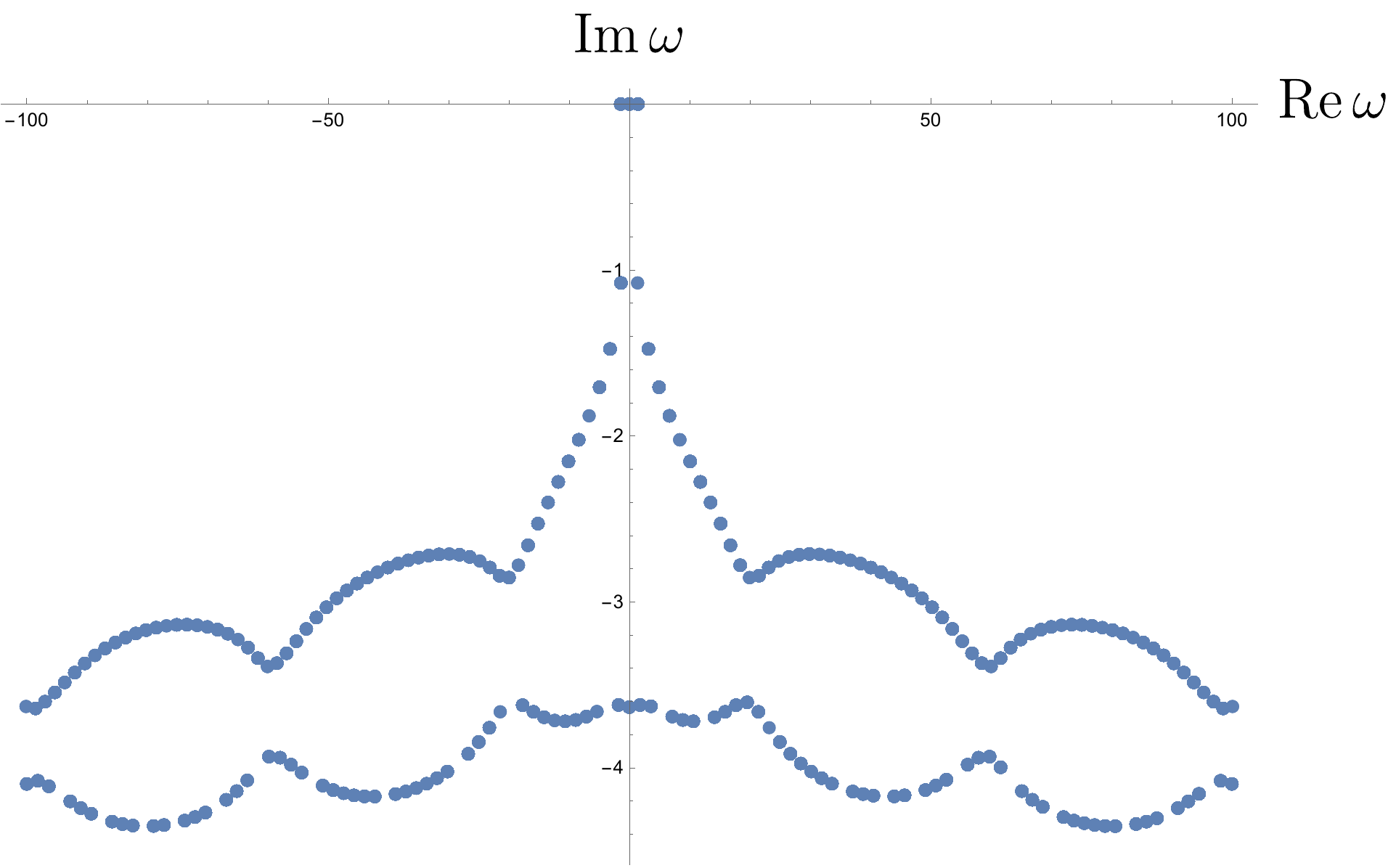}
         \caption{}
         \label{Fi:poles1}
    \end{subfigure}
    \hfill
    \begin{subfigure}[b]{0.46\textwidth}
         \centering
         \includegraphics[width=0.95\linewidth]{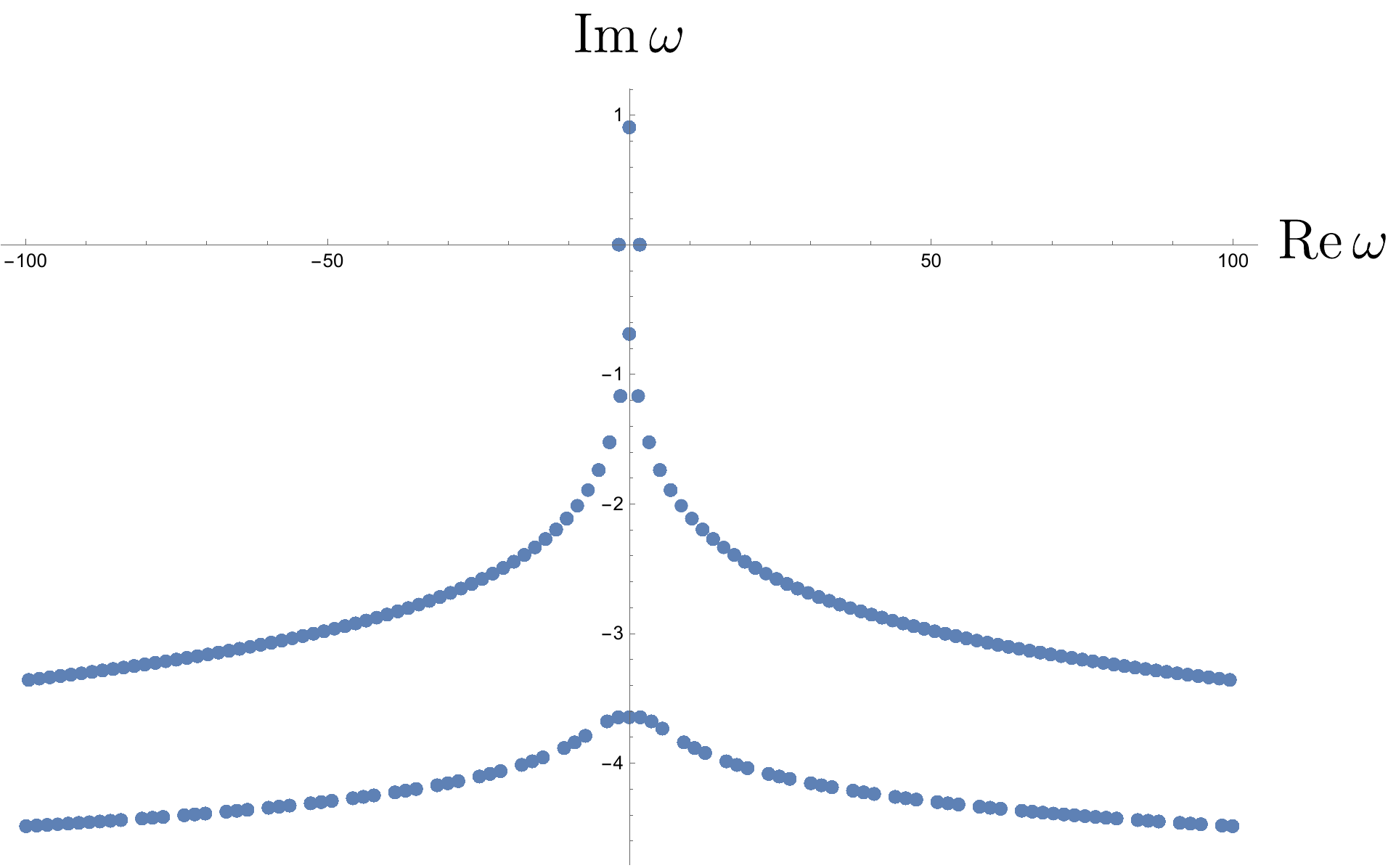}
         \caption{}
         \label{Fi:poles2}
    \end{subfigure}
    \hfill
    \caption{The distributions of the poles for a system described by Eqs.~\eqref{E:dkbfo1} and \eqref{E:dkbfo2} in the complex frequency domain. In Fig.~\ref{Fi:poles}-(a), the parameters $z_2=1.8806$ is chosen. All poles have negative imaginary parts, so the corresponding time evolution decays with time. By comparison, in Fig.~\ref{Fi:poles}-(b) $z_2=1.8$, and three of the poles have positive imaginary parts. They give runaway behavior. The rest of the parameters used in the plots are $z_1=1.8$, $\rho=0.05$, $\gamma=0.05$, $m=1$ in units of the atomic transition frequency $\omega_{\textsc{p}}$ or its inverse.} 
    \label{Fi:poles}
\end{figure}

In Fig.~\ref{Fi:poles}, we show the distributions of poles for Eqs.~\eqref{E:dkbfo1} and \eqref{E:dkbfo2} in the complex frequency plane for two different parameter sets. In Fig.~\ref{Fi:poles}-(a), all poles are located in the lower half complex frequency plane. Their negative imaginary parts lead to damping behavior in $\chi_1(t)$ and $\chi_2(t)$, and the system is expected to have asymptotic steady states. In contrast, Fig.~\ref{Fi:poles}-(b) shows that some poles have positive imaginary parts, resulting in runaway solutions where $\chi_1(t)$ and $\chi_2(t)$ grow exponentially in time. In this case, they cannot equilibrate. As we shall see later from the numerical calculations, dynamical instability tends to arise when the two atoms are sufficiently close to each other. It follows from the discussions in Appendix~\ref{S:ebedfd} that physically,  instability can be understood as the consequences of very strong Coulomb- or Lienard-Wiechert-like interaction between the internal degrees of freedom of the two atoms when they are placed at short separations, akin to the electronic dissociation of an atom in a very strong electric field.

The region of dynamical instability will thus be excluded from our evaluations of the elements of the covariance matrix of the system. In addition, in Appendix~\ref{S:ebedfd}, we have argued that when the atom is positioned next to the boundary, the Dirichlet boundary condition implies that the nonlocal contribution $\displaystyle\int^t_0\!ds\;G_{\textsc{r},\textsc{m}}(\bm{x},t;\bm{x},s)\,\hat{\chi}(s)$, arising from the presence of the boundary, can partially cancel the counterpart from $\displaystyle\int^t_0\!ds\;G_{\textsc{r},0}(\bm{x},t;\bm{x},s)\,\hat{\chi}(s)$. As a result, the effective damping constant can be substantially reduced, resulting in a much slower relaxation process~\cite{HL08}.

Here it is a good place to clarify the difference in meaning in the term ``strong/weak atom-field \textit{interaction}'' versus ``strong/weak atom-field \textit{coupling}'' in the present setting. The interaction term in the Lagrangian involves both the coupling constant $e$ and the field strength $\phi$. Accordingly, a strong atom-field interaction may result either from finite, nonvanishing values of the coupling constant, or from a large field strength. For linear field dynamics, the fluctuation-dissipation relation, for example,
\begin{equation*}
    G^{(\phi_h)}_{\textsc{h}}(\bm{x}_i,\bm{x}_j,\omega)=\coth\frac{\beta\omega}{2}\,\operatorname{Im}G^{(\phi_h)}_{\textsc{r}}(\bm{x}_i,\bm{x}_j,\omega)\,,
\end{equation*}
holds independently of the coupling constant $e$. Nonetheless, the proportionality factor equating both kernels is determined by the properties of the initial state of the field, such as the inverse temperature $\beta$, and encodes the information that governs the asymptotic dynamics of the linear system that couples with the field. This relation remains valid for any pair of spatial positions, $\bm{x}_i$, $\bm{x}_j$, although the numerical values of the kernels on both sides of the relation do depend on field amplitude. These two factors have distinct effects on the late-time entanglement dynamics between two atoms mediated by the field. The field, on one hand, correlates the internal dynamics of the two spatially separated atoms, an effect that persists even within a classical framework. On the other hand, the quantum noise arising from field fluctuations plays a more subtle role. It has been well established that field fluctuation noises can either degrade or induce entanglement between quantum systems. While this may initially appear counterintuitive, we will revisit this phenomenon in greater detail when we come to interpreting the results of atomic entanglement.

\section{Entanglement Domain}
{In this section we will show that a three-dimensional spatial region can be identified where entanglement between two atoms exist but beyond which it does not.  We call this the \textit{entanglement domain}. In this Gaussian system under study we shall use   negativity as a measure of entanglement, a summary description of which can be found in Appendix B. Of special relevance is the smaller,  called  $\lambda_-$, of the pair of symplectic eigenvalues of the partially transposed covariance matrix.   In Appendix A, we sketch the derivations of} the covariance matrix formed by the canonical pairs of the internal degrees of freedom of the two harmonic atoms (more details can be found in Refs~\cite{HHPRD16,HHPRD18}). Here we shall calculate $\lambda_-$  {for both atoms} at different spatial locations, in the vicinity of a perfectly conducting planar boundary. Both atoms get entangled via their internal degrees of freedom, coupled to a common quantum field modified by the presence of the vacuum-conductor interface, as illustrated in Fig.~\ref{Fi:afi}. 

\begin{figure}[H]
    \centering
    \hfill
    \begin{subfigure}[b]{0.17\textwidth}
         \centering
         \includegraphics[width=0.95\linewidth]{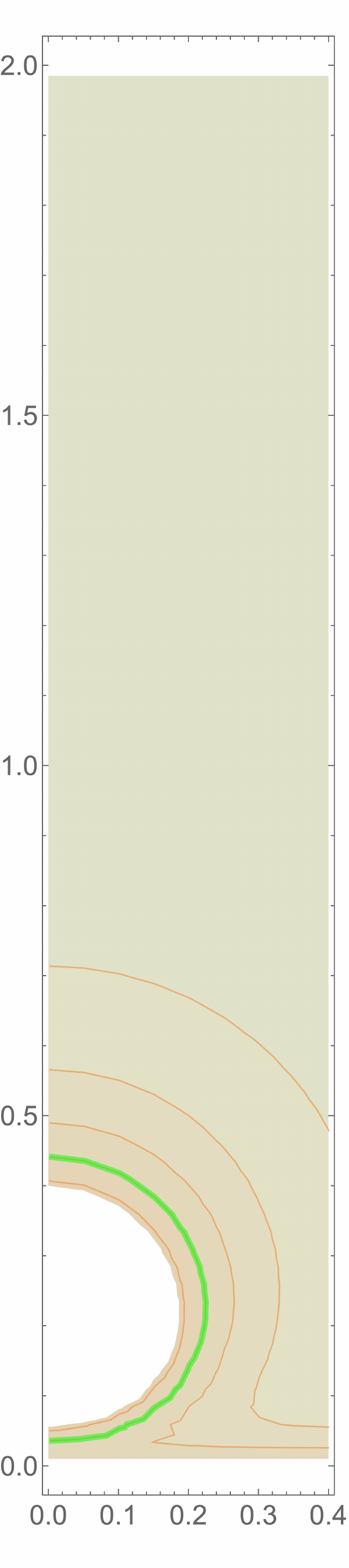}
         \caption{}
         \label{Fi:g005s02}
    \end{subfigure}
    \hfill
    \begin{subfigure}[b]{0.17\textwidth}
         \centering
         \includegraphics[width=0.95\linewidth]{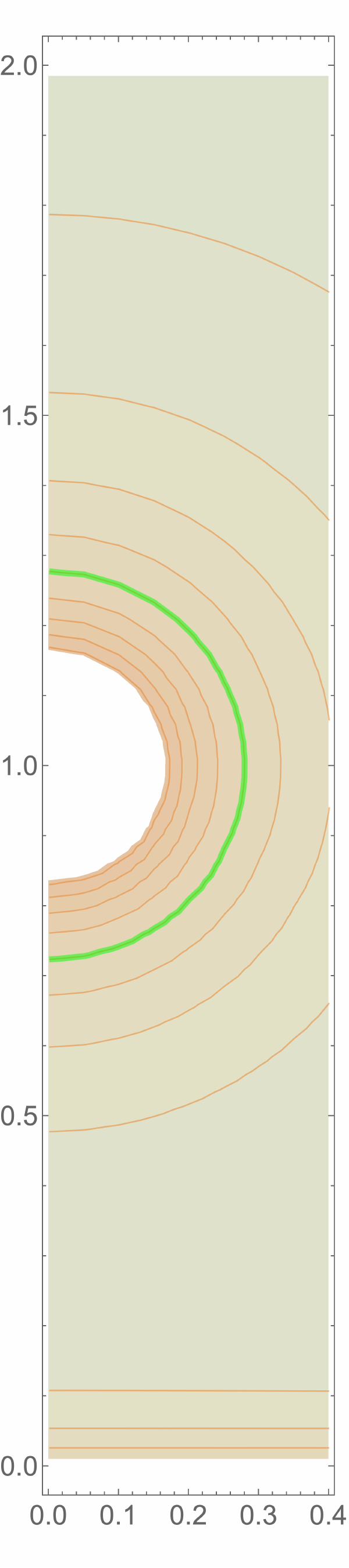}
         \caption{}
         \label{Fi:g005s10}
    \end{subfigure}
    \hfill
    \begin{subfigure}[b]{0.17\textwidth}
         \centering
         \includegraphics[width=0.95\linewidth]{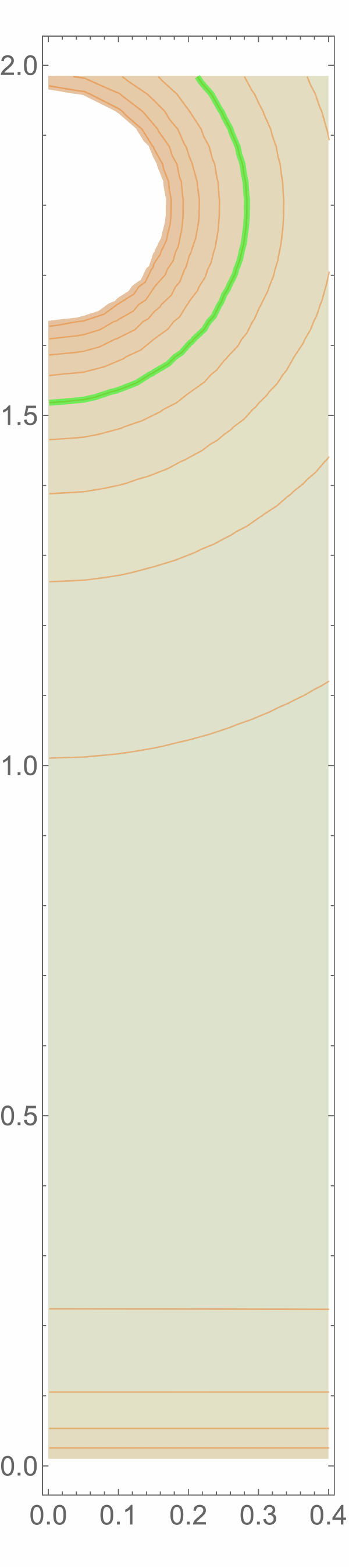}
         \caption{}
         \label{Fi:g005s18}
    \end{subfigure}
    \hfill
    \begin{subfigure}[b]{0.17\textwidth}
         \centering
         \includegraphics[width=0.95\linewidth]{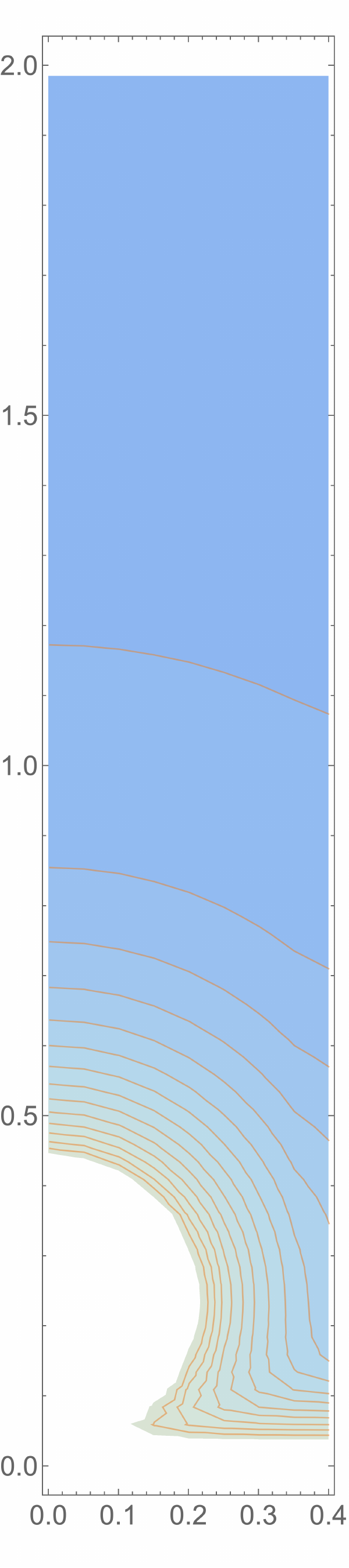}
         \caption{}
         \label{Fi:g050s02}
    \end{subfigure}
    \hfill
    \begin{subfigure}[b]{0.17\textwidth}
         \centering
         \includegraphics[width=0.95\linewidth]{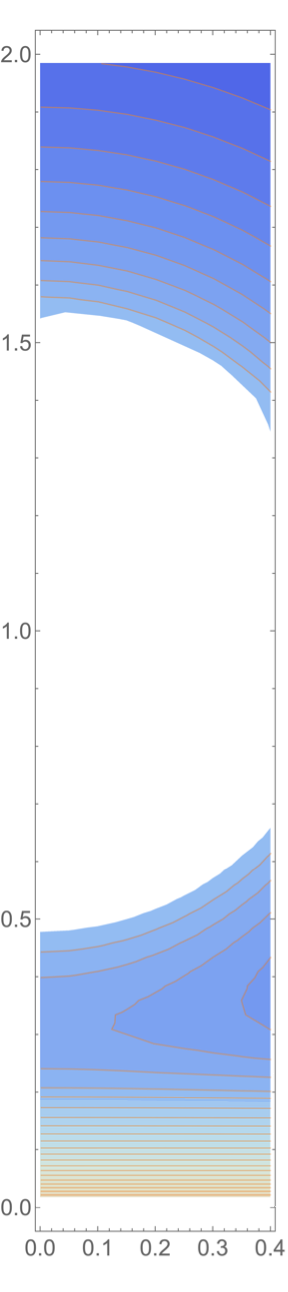}
         \caption{}
         \label{Fi:g050s10}
    \end{subfigure}
    \hfill
    \begin{subfigure}[b]{0.039\textwidth}
         \centering
         \includegraphics[width=0.95\linewidth]{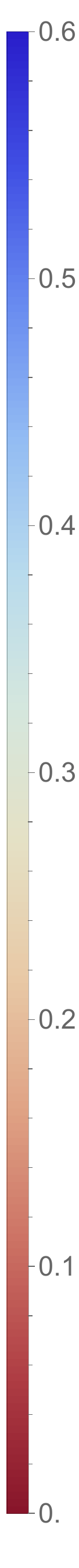}
         \label{Fi:legends}
    \end{subfigure}
    \hfill
    \caption{A density plot for $\lambda_-^2$, equivalent to negativity. When $\lambda_-^2<1/4$, two atoms are entangled. The vertical axis represents the distance $z$ to the conductor, while the horizontal axis denotes the separation between two atom. The shading tells the magnitude $\lambda_-^2$ and the green curve indicates the case $\lambda_-^2=1/4$. The first three plots corresponds to the weak coupling case $\gamma=0.05$ when one of the atoms is placed at $z_2=0.2$, $1.0$ or $1.8$ (in units of the wavelength of the transition energy), and the last two plots are results of strong coupling $\gamma=0.5$. In the strong coupling regime, basically no entanglement is possible. The white regions are dynamically unstable. The entanglement measure $\lambda^2_-$ has the same values along the thin orange contour curves.} 
    \label{Fi:sq}
\end{figure}

We are particularly interested in the values of $\lambda_-$ in the late-time steady state where both atom's internal degrees of freedom are fully relaxed. The results are presented in Fig.~\ref{Fi:sq}.  {It is worth taking a closer look while we explain the particular features.}

In Fig.~\ref{Fi:sq} we use different shades to represent the late-time numerical values of $\lambda_-^2$. The two atoms are entangled when $\lambda_-^2<1/4$, and disentangled when $\lambda_-^2>1/4$. The contour corresponding to the threshold $\lambda_-^2=1/4$ is highlighted by the thick green curves in each subplot. The vertical axis denotes the vertical distance $z_2$ of atom 2 from the perfectly conducting plate, and the horizontal axis indicates the horizontal separation $\rho$ between two atoms (Refer to Fig.~\ref{Fi:afi}). Figs.~\ref{Fi:sq}(a)--(c), display the results in the limit of weak atom-field coupling constant, with atom 1 placed at three distinct vertical distances to the boundary, $z_1=0.2$, $z_1=1$ and $z_2=1.8$ in units of the inverse of the atomic transition frequency $\omega_{\textsc{p}}$. For comparison, Figs.~\ref{Fi:sq}(d)--(e) show the results of finite coupling strength with $z_1=0.2$ and $z_1=1$, respectively.

The white circular regions correspond to parameter values for which the internal degrees of freedom of two atoms exhibit dynamical instability. This phenomenon is consistent with the interpretation that such instability arises from the large field amplitude for a Coulomb-like or Lienard-Wiechert-like interaction when both atoms are very close to each other. Comparing the extent of these instability regions in Figs.~\ref{Fi:sq}(a)--(c) and Figs.~\ref{Fi:sq}(d)--(e), we observe that their overall size generally increases with the coupling strength $e$, except in cases where both atoms are located adjacent to the boundary. In these latter cases, the boundary effects manifest as noticeable distortions of the contour curves away from circular (or spherical, considering the cylindrical symmetry) shapes, since $\lambda^2_-$ takes on the same values along each contour.

The noteworthy feature in Figs.~\ref{Fi:sq}(a)--(c) which enables the definition of \textit{entanglement domain }is that the contour corresponding to $\lambda_-^2=1/4$ encloses a nearly spherical region.   If both atoms are placed inside this domain, they can become entangled at late times, given some suitable form of atom-field interaction and for a prescribed initial state of the field. Crucially, this entanglement is independent of the initial states in the dynamics of the atoms' internal  degrees of freedom. This follows from the fact that the asymptotic equilibrium state of a linear Gaussian reduced system is insensitive to its initial configuration. Its late-time properties are instead governed entirely by the environment, taking the perspective of open quantum systems. The notion of the entanglement domain is meaningful only when the internal degrees of freedom of both atoms have been fully relaxed to their respective asymptotic steady state. In this regime, the relevant covariance matrix elements that define the entanglement measure stabilize to constants, yielding an \textit{unambiguous radius of influence} in the entanglement between the atoms. In contrast, during the transient regime, the entanglement measure is a function of time, and typically exhibits rapid oscillations due to fast field dynamics. These temporal fluctuations obscure the boundary of the entanglement domain, rendering it ill-defined in the non-equilibrium phase.

\begin{figure}
    \centering
    \includegraphics[width=0.5\linewidth]{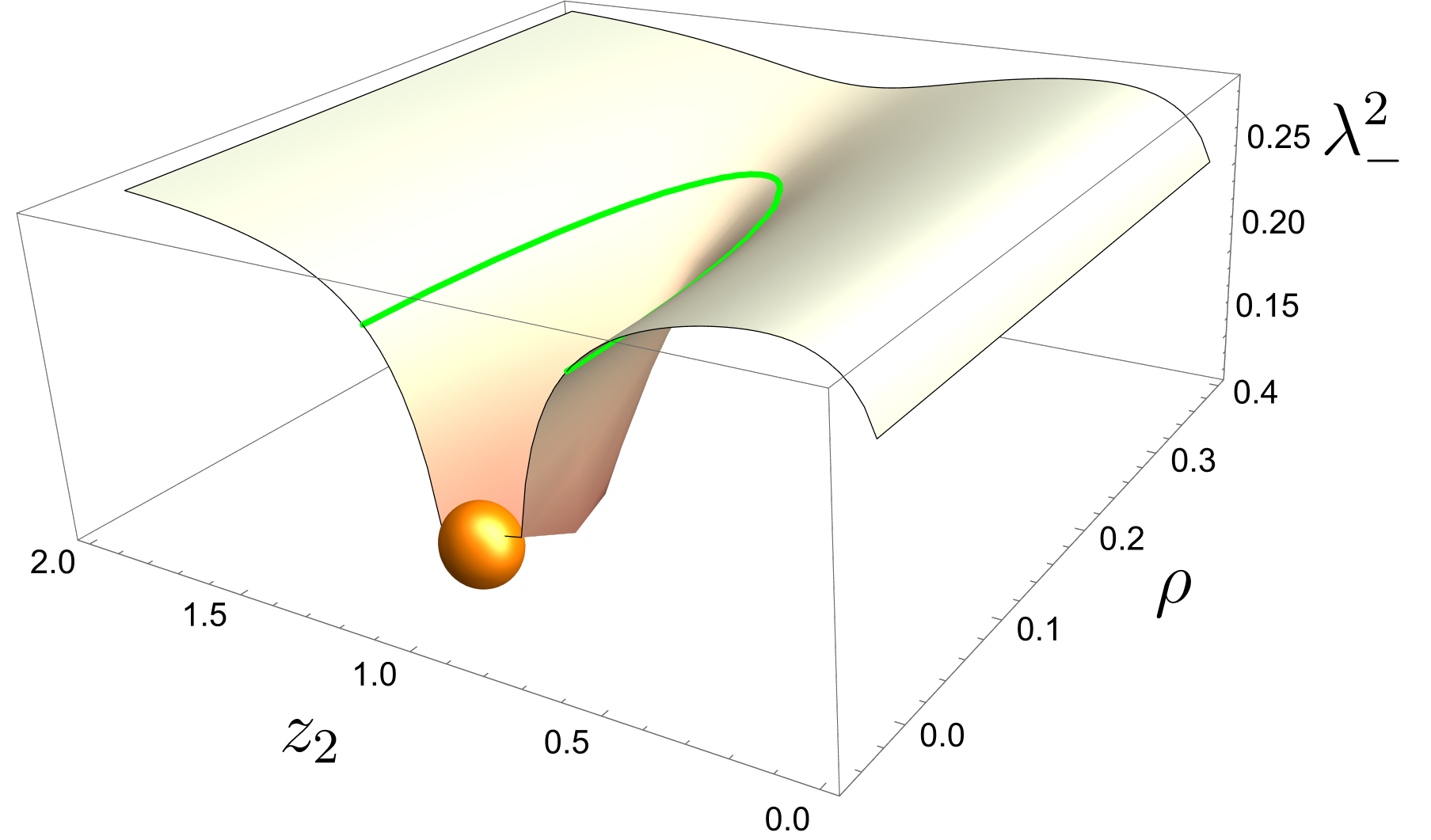}
    \caption{This is an alternative representation of Fig.~\ref{Fi:sq}(b). The sphere denotes the location of atom 1. The $(x,y)$ coordinates of any point on the surface denote the location of the second atom, relative to atom 1, while the $z$ coordinate gives the extent of entanglement between two atoms. Again the thick green curve is the boundary between entanglement and disentanglement.}
    \label{Fi:g005s103D}
\end{figure}

\section{Dependence of Entanglement on Interaction strength and Spatial Separation}

In this section we give details on the dependence of two-atom entanglement mediated by a quantum field in the presence of a conductor surface  on various parameters involved.  But, before that, some comments on the role of the quantum field shared by both atoms in mediating their entanglement are useful.   

\subsection{Quantum field fluctuations have correlations}

It is a well-known yet somewhat enigmatic fact that a quantum field can either generate or suppress entanglement in a bipartite quantum system. Creation of entanglement by the environment quantum field comes somewhat as a surprise if one holds the view that field fluctuations acting as a source of quantum noise, tend to destroy phase coherence between subsystems, and thus degrade their correlation. {The root cause of the fallacy in this view is ignoring the fact that quantum fields are intrinsically coherent and quantum fluctuations should not be regarded as random and identified by stochastic variables; instead, they possess intrinsic spatial and temporal correlations~\cite{HH19,BHH23}. Consequently, these correlations eventually imprint themselves onto the late-time dynamics in the components of a linear system. As emphasized earlier and  {sketched in Appendix A (Details are provided in Ref.~\cite{HHPRD18})}, the relaxation behavior of a linear system is predominantly governed by the properties of its environment, rather than its initial state.}

This behavior is evident in the expressions of the late-time covariance matrix elements associated with the internal degrees of freedom of both atoms, as   discussed in Appendix A. Therefore, even when starting from separable initial conditions, the system may evolve under the influence of a correlated environment into an entangled steady state.  It is important to recognize that entanglement creation or destruction between the two atoms is determined in relation to their initial settings. Since the asymptotic value of the entanglement measure $\lambda_-^2$ is independent of the initial states of the atoms' internal dynamics, its comparison with the corresponding initial value determines whether entanglement has been generated or destroyed during the nonequilibrium evolution. In this setup, the long-time behavior of $\lambda_-^2$ depends solely on the initial state of the field, the atom-field coupling strength, and the spatial configuration of the atoms.

\subsection{Interaction strength dependence}

As remarked earlier, the influence of quantum environmental field on entanglement arises from the integrated effect of the atom-field coupling strength and the amplitude of field fluctuations, the latter can be represented for Gaussian systems as a form of noise. The emergence of an entanglement domain then indicates that when both atoms are sufficiently close to each other, their mutual interaction mediated by the field becomes strong enough to overcome the decohering effect of the noise in the field. {This enhanced field-induced coherence will sustain their entanglement.}  Negativity being a quantifiable measure, we can see from Figs.~\ref{Fi:sq}(a)--(c) that the degree of entanglement improves with closer separations between the atoms, until dynamical instability sets in. {Simply put, considering only the interaction strength factor, quantum entanglement increases when the atoms get closer, which is not unexpected.}

A similar behavior is also observed in the model studied in Ref.~\cite{HHPRD16}, which involves two spatially separated Unruh-DeWitt detectors, with internal degrees of freedom modeled by harmonic oscillators, same as the harmonic atoms considered here. The two detectors are placed in flat space without boundary and they interact not only via a shared quantum field (called indirect interaction) but also through direct coupling. The direct interaction is bilinear in the internal degrees of freedom of both detectors and is independent of the distance between the detectors.  We can see from Fig.~1 of Ref.~\cite{HHPRD16} that when the two detectors get sufficiently close, the entanglement between them emerges and is strengthened as their separation $\ell$ decreases. The major difference from the present case is that,  {since the direct interaction in that case is independent of mutual separation, it can overshadow the field noise and thus keep the detectors} entangled even at large separations.

The same spatial dependence of $\lambda^2_-$ is also evident in Figs.~\ref{Fi:sq}(d)--(e). The numerical values of  $\lambda^2_-$ still decrease with shorter separation. However, in these cases, the atoms remain separable across all dynamically stable configurations, and no entanglement is observed. The large atom-field coupling is the culprit. It amplifies the overall amplitude of free field fluctuations. As a result, the vacuum noise from the field becomes too disruptive for the atom-atom interaction to sustain the mutual coherence between atoms needed for entanglement to emerge.

\subsection{Boundary distortions of entanglement domain}

Just as the instability region becomes distorted when both atoms are placed near the boundary, similar distortions in the shape of the entanglement domain are observed in Figs.~\ref{Fi:sq}, highlighting the influence of the perfectly conducting boundary. To elucidate these effects, we first observe that the values of $\lambda^2_-$ also decrease when atom 2 approaches the boundary, and the contour curves representing constant $\lambda^2_-$ gradually transition from circular to horizontal, provided that atom 1 is not too close to the conducting plate. As illustrated in Figs.~\ref{Fi:sq}(b), (c) and (e), the values of $\lambda^2_-$ become primarily dependent on the vertical distance of atom 2 to the boundary, with minimal sensitivity to the horizontal separation.

This trend can be attributed to two factors. First, the mutual interaction between atoms weakens with increasing separation. Second, and more significantly, the presence of the boundary substantially suppresses the interaction. Such suppression can be understood by decomposing the interaction between atom 1 and atom 2 in the presence of the boundary into two components: the direct interaction between the atoms and the virtual interaction between atom 1 and the mirror image of atom 2, as inferred from the method of images used in constructing the retarded Green's function $G^{(\phi_h)}_{\textsc{r}}(x,x')$. Since the mirror image of atom 2 carries the opposite polarity under the Dirichlet condition, these two contributions tend to cancel each other, resulting in diminished interaction strength.

In contrast, the field fluctuations locally acting on atom 2 depend only on the vertical distance from the conducting plate, and are largely independent of the horizontal separation between atom 1 and atom 2. This spatial anisotropy accounts for the emergence of horizontal features of contour lines.  Furthermore, as atom 2 approaches the boundary, the imposed Dirichlet condition reduces the amplitude of field fluctuations, thereby weakening their decohering influence. This suppression explains the observed decrease in $\lambda^2_-$ near the boundary.

The preceding analysis may suggest that quantum entanglement between atoms is more favorably maintained when both atoms are placed closer to the boundary, especially given that the contours extend to large values of $\rho$ when both $z_1$ and $z_2$ are small. However, a closer inspection of Fig.~\ref{Fi:sq}(a) reveals a more nuanced picture.  As atom 2 approaches the plate, while atom 1 remains fixed near the conducting plate, the green entanglement contour which defines the entanglement domain is noticeably compressed toward the white region that corresponds to dynamical instability. This implies that entanglement exists only within a narrow annular region, where the mutual interaction between atoms is strong enough to generate entanglement at late times, yet still below the instability threshold.

This behavior indicates that near the plate, the atoms must be positioned extremely close to each other in the horizontal direction in order for entanglement to survive at late times. Although local vacuum fluctuations of the field on each atom are suppressed by the boundary condition, the mutual interaction between them appears to be more significantly attenuated. This suppression likely results from the enhanced cancellation between direct and virtual-image mediated field contributions, as both atoms and their mirror images interact destructively. As a consequence, the spatial region harboring entanglement at late times contracts as atom 2 approaches the plate. This suggests that the overall boundary influence does not seem to favor entanglement between atoms. Indeed, according to Figs.~\ref{Fi:sq}(a) and (d), the contour curves of constant $\lambda^2_-$ exhibit sharp turns near the lower left corner of each plot. These sharp transitions signal a shift in the dominant influence, from mutual interaction to field noise, reflecting the competition between these two effects.

Nonetheless, this reasoning should be interpreted with some caution. As noted at the end of Sec.~\ref{S:verued}, the treatment of reduced dynamics for quantum open systems in extreme proximity to a perfectly conducting boundary is quite involved. {The required renormalization procedure has ambiguities which affects its physical interpretation, and the boundary condition itself is a highly idealized abstraction.  A definitive assessment of extreme proximity boundary effects on entanglement probably needs not only a detailed theoretical treatment as we try to provide here albeit with simple model studies, but also knowledge of the material properties of the actual boundary surface. }

\subsection{Crossing between entangled and separable, in space and in time}
Let us examine the formation of entanglement domain in time and the conditions for its disappearance. It is known that entanglement between two quantum subsystems vanishes completely in a finite time, a phenomenon known as entanglement sudden death~\cite{DodHal,YuEberly1,Almeida,YuEberly2}. Compare with most other quantum information relevant quantities, such as coherence or purity, their qualitative changes usually come  gradually, not sudden. This characteristic is particularly striking given that the interaction mediated by the massless linear quantum field has infinite range.  Instead of a sudden change of entanglement in time, here, we have a rather sharp demarcation in space,  which enables the definition of an entanglement domain: Crossing from a region inside the entanglement domain to a region outside it is rather abrupt. One can regard this as the spatial analog of sudden death.  What is equally interesting is to ask about the temporal behavior of entanglement, such as the formation of an entanglement domain \footnote{Before the internal dynamics is fully relaxed and the system reaches a steady state, we cannot rule out the appearance and disappearance of entanglement, maybe even multiple times,  and the size of the entanglement domains, if they can be identified  at those times, which may also change with time. Only a full knowledge of the nonequilibrium evolution of the idf of the atoms interacting with their common boundary-altered quantum field can tell.}

{Because our theory can treat the nonequilibrium dynamics of the atoms' internal degrees of freedom, we should  {in principle} be able to answer these questions.  For Gaussian states, the answers come  easily because the threshold between separable and entangled states is sharply defined by $\lambda_-^2=1/4$, while $\lambda_-^2$ can assume any nonnegative values. Hence, following the evolution of $\lambda_-^2$ in time will show any  abrupt transitions in entanglement, such as sudden death or revival. They occur precisely when the curve representing $\lambda_-^2$ crosses this threshold. }

\begin{figure}
    \centering
    \hfill
    \begin{subfigure}[b]{0.46\textwidth}
         \centering
         \includegraphics[width=0.95\linewidth]{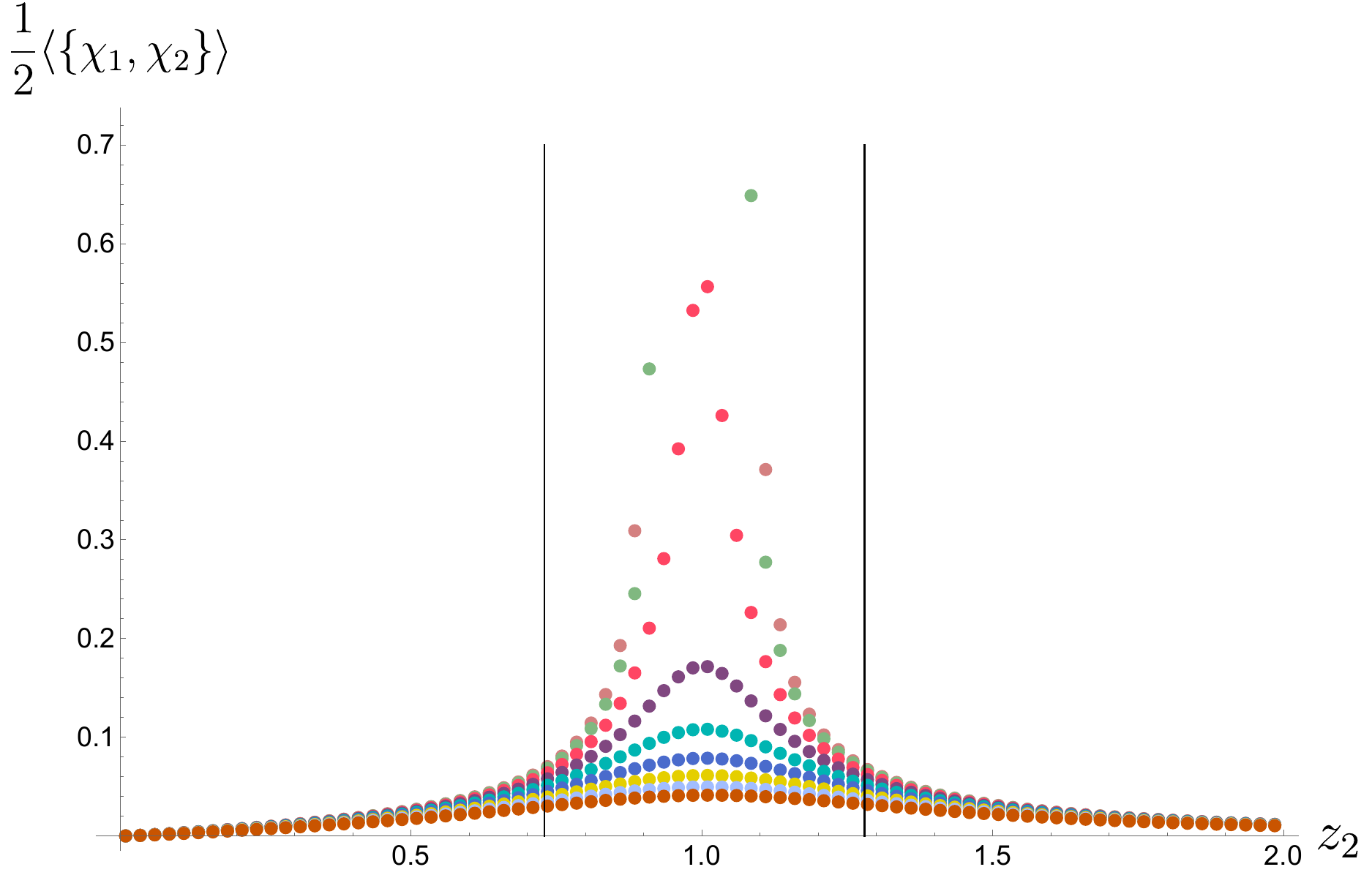}
         \caption{}
         \label{Fi:x1x2}
    \end{subfigure}
    \hfill
    \begin{subfigure}[b]{0.46\textwidth}
         \centering
         \includegraphics[width=0.95\linewidth]{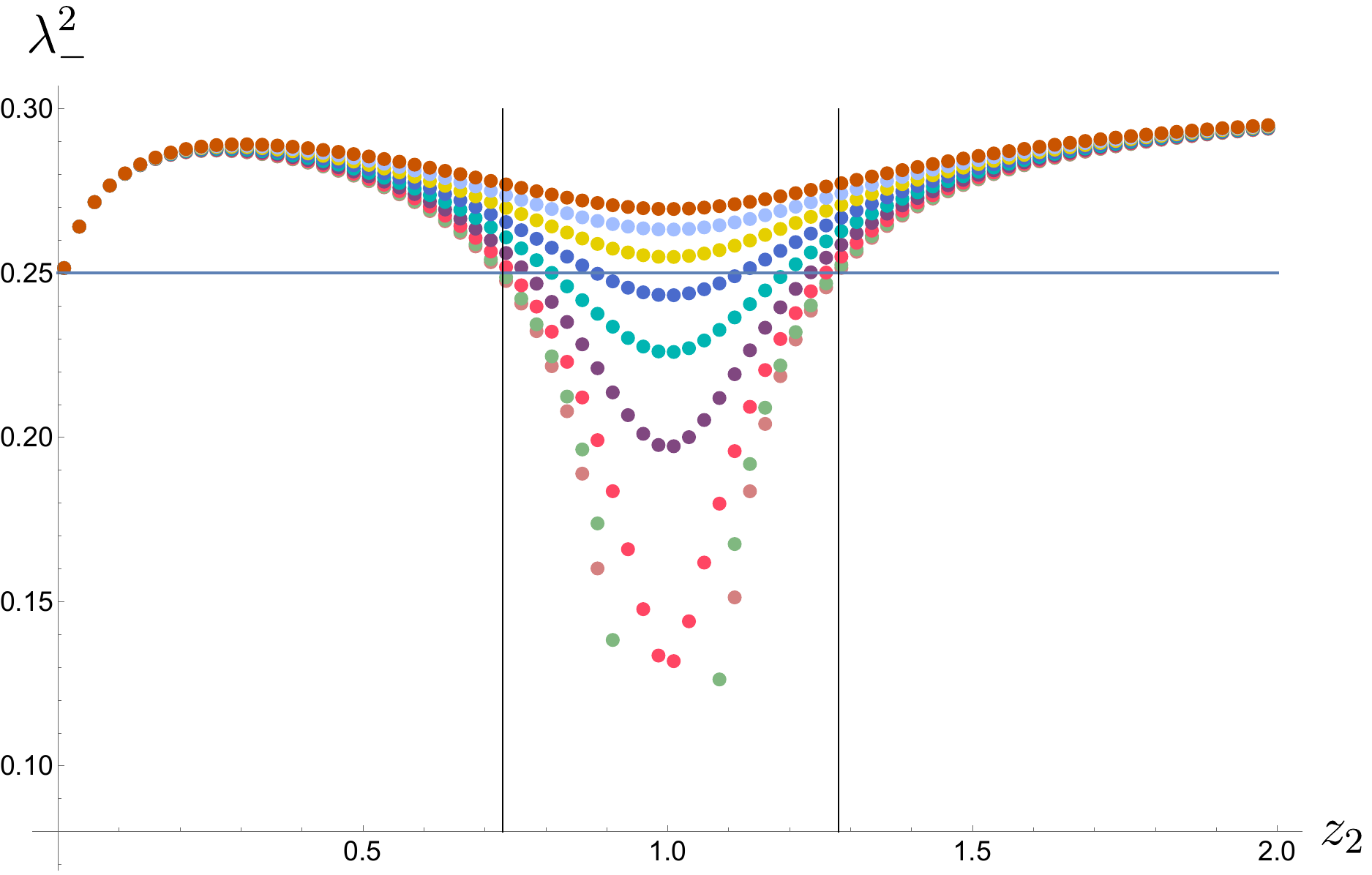}
         \caption{}
         \label{Fi:lambda}
    \end{subfigure}
    \hfill
    \caption{(a) Correlation between the two atoms when atom 1 is fixed at $z_1 = 1$, and atom 2 is placed at various vertical positions $z_2 \in [0.01, 2]$. Each curve corresponds to a different horizontal separation $\rho$. (b) Corresponding values of $\lambda_-^2$; the horizontal line at $1/4$ marks the entanglement threshold. Entanglement exists in the region where $\lambda_-^2 < 1/4$. When entanglement is present, the correlation between atoms is typically strong, reflecting enhanced mutual interaction mediated by the ambient field.} 
    \label{Fi:correlation}
\end{figure}

Here, it is also illustrative to highlight the role of mutual correlation between two atoms in their entanglement. Fig.~\ref{Fi:correlation}(a) depicts the correlation between the two atoms, while Fig.~\ref{Fi:correlation}(b) identifies the spatial interval where entanglement occurs. The region between the pair of vertical lines in Fig.~\ref{Fi:correlation}(b) corresponds to configurations where $\lambda_-^2 < 1/4$, indicating entanglement. The results suggest that strong mutual correlation is a necessary, but not sufficient, condition for entanglement: without sufficient correlation, entanglement does not arise.

\begin{figure}
    \centering
    \hfill
    \begin{subfigure}[b]{0.46\textwidth}
         \centering
         \includegraphics[width=0.95\linewidth]{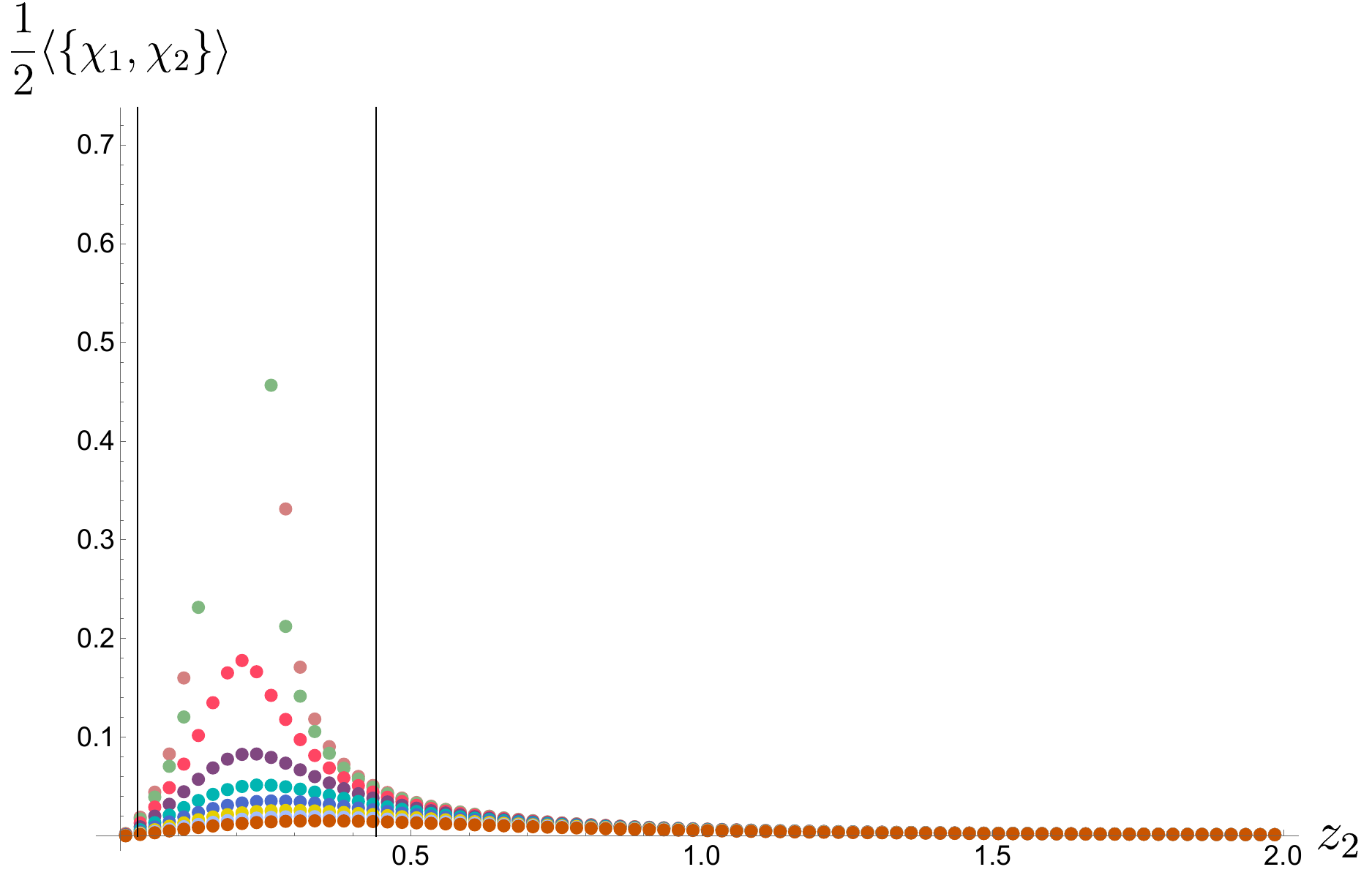}
         \caption{}
         \label{Fi:x1x2z02}
    \end{subfigure}
    \hfill
    \begin{subfigure}[b]{0.46\textwidth}
         \centering
         \includegraphics[width=0.95\linewidth]{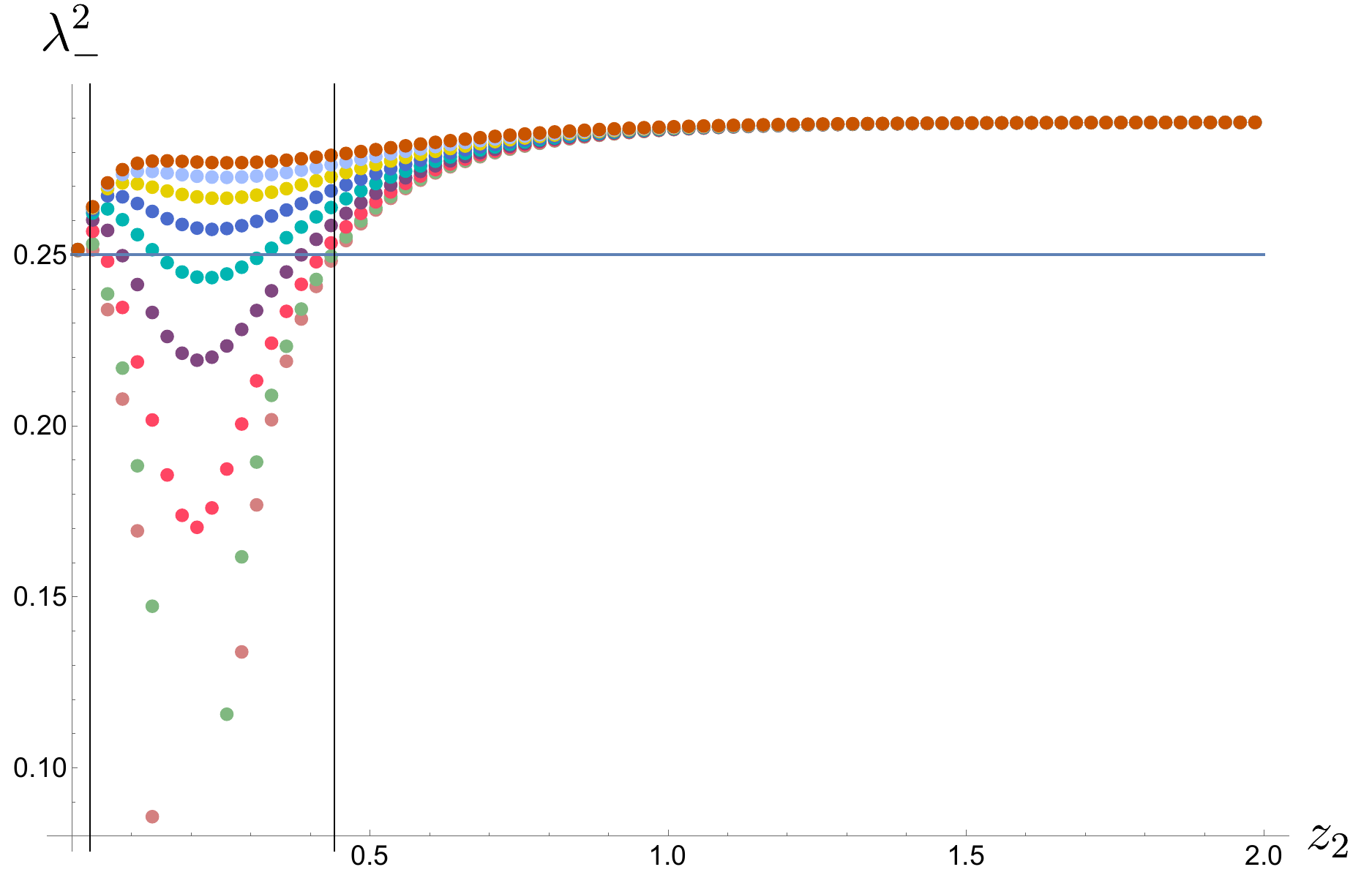}
         \caption{}
         \label{Fi:lambda02}
    \end{subfigure}
    \hfill
    \caption{Correlation between the two atoms when atom 1 is fixed at $z_1 = 0.2$, and atom 2 is placed at various vertical positions $z_2 \in [0.01, 2]$. The results suggest that the presence of the boundary condition tends to suppress the correlation between the atoms, as evidenced by the narrower intervals and less pronounced peaks of the correlation curves (see also Figs.~\ref{Fi:sq}(a)--(c)). Nevertheless, the overall influence of the boundary remains relatively modest. The primary factor governing the emergence of entanglement between the atoms is their mutual interaction, except perhaps in scenarios where both atoms are situated in extremely close proximity to the boundary.} 
    \label{Fi:correlationz02}
\end{figure}

A similar conclusion can be drawn from Fig.~\ref{Fi:correlationz02}, which considers the case where atom 1 is placed close to the conducting plate. Even in the presence of boundary-induced modifications, the strength of mutual correlation remains a key factor in determining whether entanglement emerges.

\subsection{Connecting with results from Paper II}\label{S:bfbieur}

In Paper II \cite{AFD2} of this series, we investigated the quantum entanglement between a single atom and the ambient quantum field occupying the half-space outside a perfectly conducting boundary. The present setup reproduces that earlier configuration when one atom, say atom 1, is moved to spatial infinity, $z_1\to\infty$. At such a distance, atom 1 neither senses the boundary nor interacts appreciably with atom 2. In fact, both atoms remain separable, as can be inferred from Fig.~\ref{Fi:sq}. Under these conditions, the entanglement between atom 2 and the boundary-modified field is essentially the same as in the single-atom case studied previously.

When the combined atom-field system is initially prepared in a pure state, two common entanglement measures are employed: purity and the von Neumann entropy of the atom (or field) alone. Under atom-field interaction, an initial pure state generally evolves into a superposition of product states.  If this final state cannot be expressed as a single product state by any linear transformation, the total system is no longer pure and the atom and field are entangled.

Purity quantifies the degree of pureness/mixedness of the reduced state of the atom as the system,  by   tracing over the field's degree of freedom as its environment (or vice versa, depending on what is needed for the problem). $\mu=1$ corresponds to a pure (disentangled) state, whereas $\mu<1$ signals entanglement.  The corresponding von Neumann entropy is zero for a pure product state and positive for an entangled state. Thus, both purity and von Neumann entropy provide complementary measures of atom-field entanglement when the global system starts in a pure state.

\begin{figure}
    \centering
    \hfill
    \begin{subfigure}[b]{0.46\textwidth}
         \centering
         \includegraphics[width=0.95\linewidth]{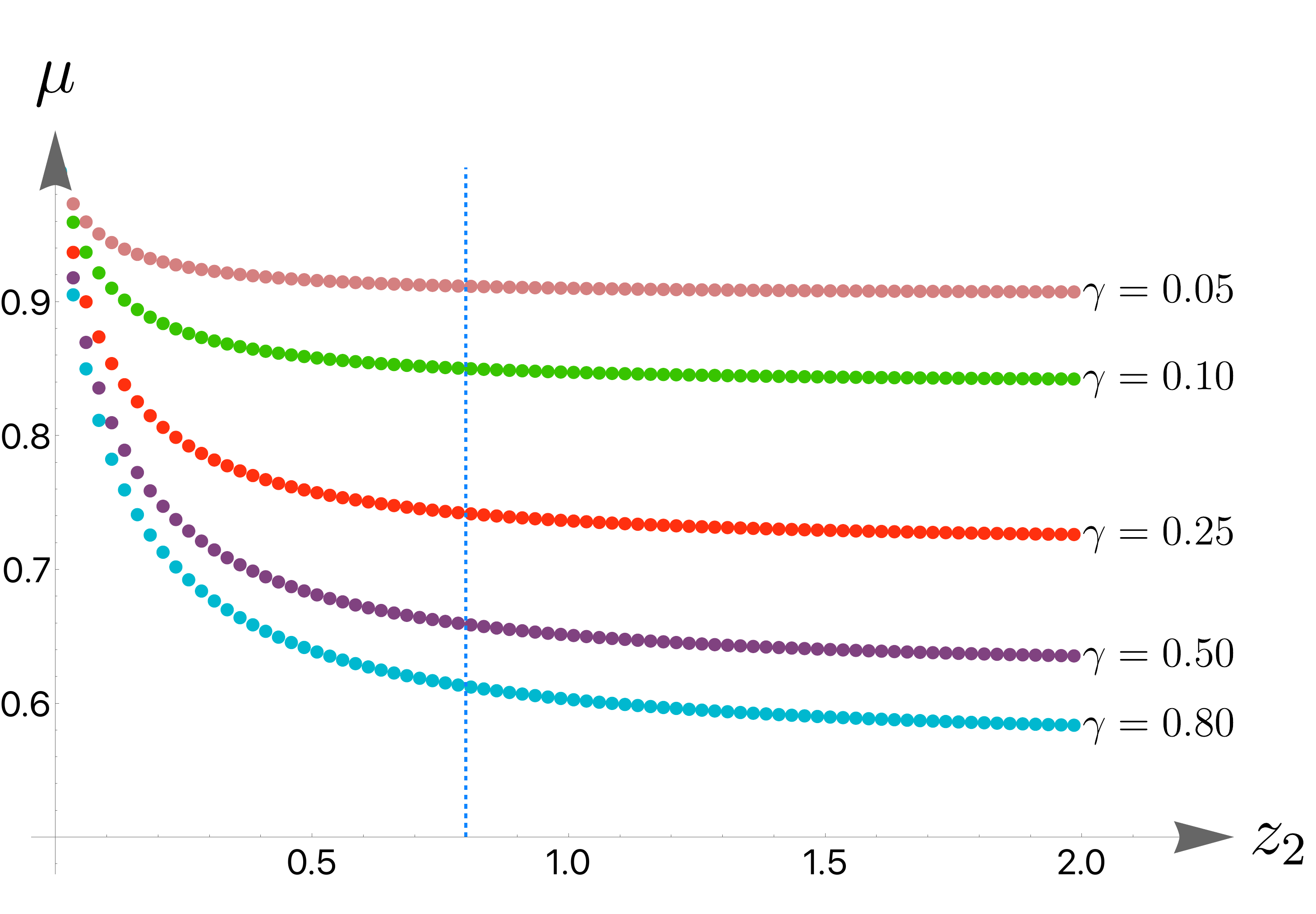}
         \caption{}
         \label{Fi:purity}
    \end{subfigure}
    \hfill
    \begin{subfigure}[b]{0.46\textwidth}
         \centering
         \includegraphics[width=0.95\linewidth]{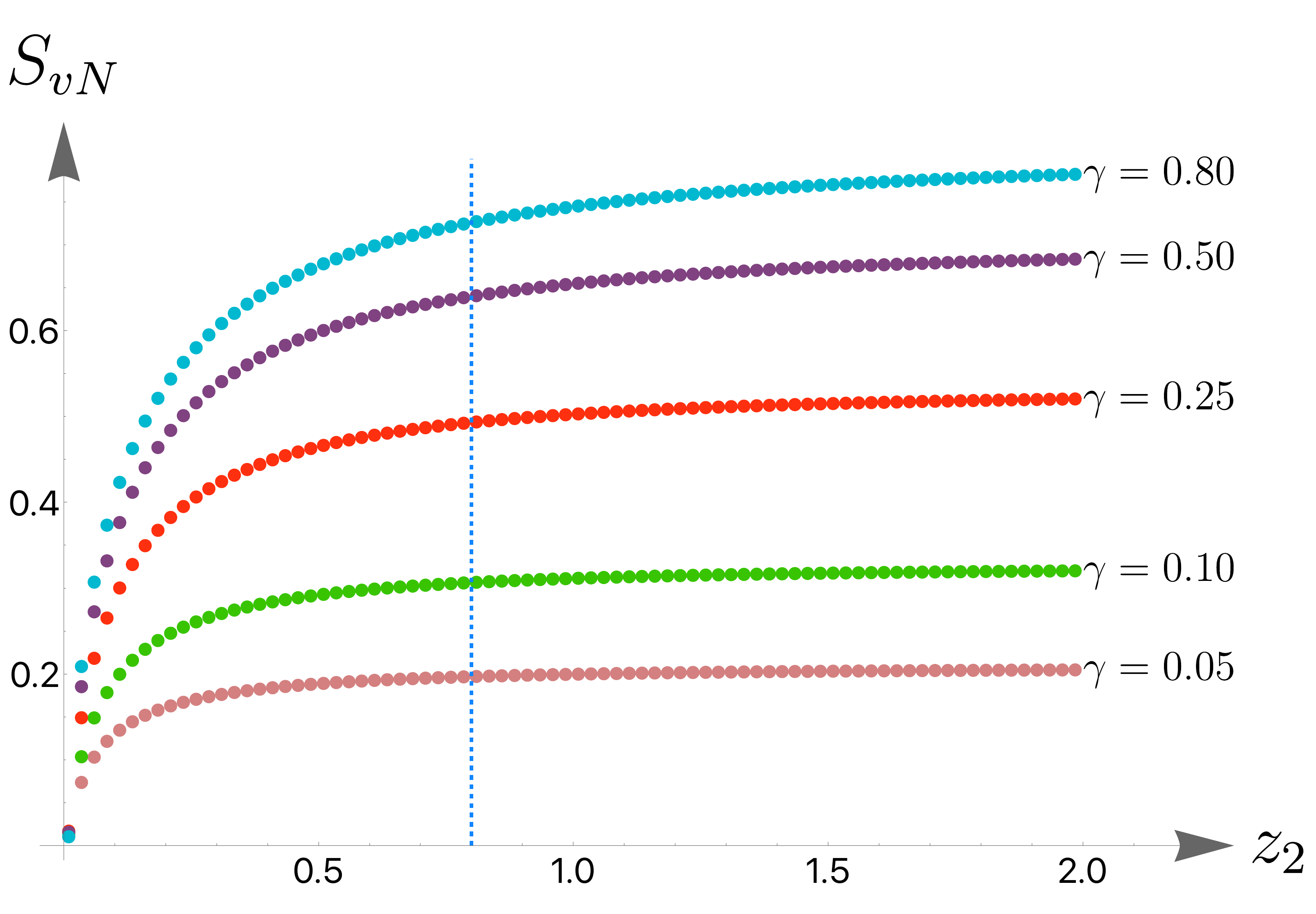}
         \caption{}
         \label{Fi:vonNeumann}
    \end{subfigure}
    \hfill
    \caption{We recreate the results in Paper II~\cite{AFD2} within the present model by placing atom 1 at $z_1=10$. The region to the left of the vertical line corresponds to the plot range in Ref.~\cite{AFD2}. The agreement between the two results is virtually exact, consistent with the expectation that atom 1 has a negligible effect on atom 2. Both purity and von Neumann entropy all indicate that the quantum entanglement between atom and quantum field degrades when atom 2 approaches the perfectly conducting boundary.} 
    \label{Fi:afentanglement}
\end{figure}

In Fig.~\ref{Fi:afentanglement}, we reproduce the results in the previous paper by placing atom 1 at $z_1=10$ (in units of the inverse of the atomic transition frequency $\omega_{\textsc{p}}$). Panel (a) plots the purity $\mu$ of atom 2 as a function of its distance $z_2$ from the conducting plate, while panel (b) provides the same information in terms of the von Neumann entropy of atom 2.

Let us consider purity as an illustrative example. For a fixed atom-field coupling strength, the purity increases as atom 2 moves closer to the boundary. This indicates that the atom-field entanglement diminishes near the perfectly conducting plate. Such behavior is consistent with earlier arguments suggesting that the effective atom-field coupling is significantly suppressed in close proximity to the boundary. In contrast, such suppression can be mitigated by increasing the damping constant $\gamma$ since it is related to the coupling constant by $\gamma=e^2/(8\pi m)$.

\section{Summary}

{In this paper, we investigate the characteristics of quantum entanglement at late-time between two neutral atoms whose internal degrees of freedom are simultaneously coupled to a common ambient quantum field whose configuration is modified by the presence of a perfectly conducting planar boundary.}

{We first identify the spatial region of dynamical instability in the dynamics of the internal degrees of freedom of the two atoms. When the interatomic separation falls below the characteristic size of this region, the system exhibits runaway behavior, signaling the onset of dynamical instability.}

{For dynamically stable configurations, once the internal dynamics of both atoms has relaxed to an asymptotic steady state, a well-defined spatial domain of entanglement, characterized by the negativity, emerges. Within this domain, the atoms remain entangled at late times, while outside it, they become separable. The size of this entanglement domain depends on the atom-field coupling strength and the initial state of the field, but not on the atoms’ initial states. Its formation reflects the competition between the nonlocal interaction mediated by the field and the local quantum noise acting on each atom.}

The presence of the boundary on the ambient field, {effected by the imposed Dirichlet condition, tends to reduce the interaction between the atoms.} For instance, an atom placed closer to the boundary exhibits a longer relaxation time in its internal dynamics. The boundary also introduces spatially anisotropic influences on the ambient field. {These influences can be understood in terms of mirror images of both atoms, in the replace of a perfectly reflecting boundary.} The numerical results indicate that the parameter $\lambda_-^2$ decreases, signaling the trend towards improved entanglement, when atoms are placed at locations closer to the boundary. However, subtleties arise when the atoms are posited in the extreme vicinity of the boundary. There, the nonlocal field-mediated interaction appears to be suppressed more strongly than the local field noise. As a result, the quantum entanglement between these two atoms is not favored,  unless the atoms are very closely placed in their lateral (horizontal) positions. This is a manifestation of the competitive interplay between the interatomic interaction and the field noise. The characteristics of entanglement in this regime may also be sensitive to model details {and boundary material properties}.

\vskip18pt
\noindent{\bf Acknowledgments} 
J.-T. Hsiang is supported by the National Science and Technology Council of Taiwan, R.O.C. under Grant No.~NSTC 113-2112-M-011-001-MY3.   B.-L. Hu enjoyed the gracious hospitality of colleagues at the Institute of Physics, Academia Sinica, Nankang and at the National Center for Theoretical Sciences at National Tsing Hua University, Hsinchu, Taiwan, R.O.C. where part of this work was carried out.

\vskip18pt
\appendix
\section{The Green's functions in a nutshell}\label{S:ebedfd}
Consider a massless scalar field $\hat{\phi}$ in three-dimensional Minkowski space confined to the half-space $z>0$, with a perfectly reflecting, infinite planar boundary imposed at $z=0$. The normalized positive-frequency mode functions $u_k$ satisfying the Dirichlet boundary condition are given by
\begin{equation}
    u_k(x_{\perp},z,t)=\sin k_zz\,e^{ik_{\perp}\cdot x_{\perp}}e^{-i\omega t}\,,
\end{equation}
where $x_{\perp}$, $k_{\perp}$ represent the components of $x$, $k$ in the plane normal to the $z$ axis, and $\omega=\sqrt{k_{\perp}^2+k_z^2}$. In terms of these mode functions, the scalar field operator $\hat{\phi}$ admits the expansion
\begin{equation}
    \hat{\phi}(x)=\int_{-\infty}^{\infty}\!\frac{d^2k_{\perp}}{2\pi}\int_0^{\infty}\!\frac{dk_z}{(2\pi)^{\frac{1}{2}}}\;\frac{2}{\sqrt{2\omega}}\,\biggl[\hat{a}_{k}\,u_k(x_{\perp},z,t)+\mathrm{h.c.}\biggr]\,.
\end{equation}
where $x$ is a position four-vector denoting $x=(\bm{x},t)$ with $\bm{x}=(x_{\perp},z)$. The Pauli-Jordan function, which represents the commutator of two field operators, is defined by
\begin{equation}
    G_{\textsc{c}}(x,x')=i\,\bigl[\hat{\phi}(x),\,\hat{\phi}(x')\bigr]\,. 
\end{equation}
In unbounded free space, this function, denoted by $G_{\textsc{c},0}(x,x')$, takes a simple form
\begin{equation}\label{E:diuer}
    G_{\textsc{c},0}(x,x')=\frac{1}{4\pi R}\int_{-\infty}^{\infty}\!\frac{d\omega}{2\pi}\;\Bigl(e^{+i\omega R}-e^{-i\omega R}\Bigr)\,e^{-i\omega\tau}\,,
\end{equation}
with $\tau=t-t'$ and $R=\lvert\bm{x}-\bm{x}'\rvert$. This integral form will be useful later.

It is then straightforward to show that, in the presence of a planar boundary, the Pauli-Jordan function $G_{\textsc{c}}(x,x')$ in the half-space can be expressed as a superposition of free-space contributions evaluated at the source point $x'$ and its image $\tilde{x}'$ with respect to the boundary
\begin{equation}\label{E:dbkier}
    G_{\textsc{c}}(x,x')=G_{\textsc{c},0}(x,x')-G_{\textsc{c},0}(x,\tilde{x}')\,,
\end{equation}
where $\tilde{x}'$ is the image of $x'$ with respect to the planar boundary at $z=0$. Sometimes, we will denote the image contribution as $G_{\textsc{c},\textsc{m}}(x,x')=-G_{\textsc{c},0}(x,\tilde{x}')$ to emphasize that it arises as a consequence of the boundary condition. With this notation, Eq.~\eqref{E:dbkier} takes the form 
\begin{equation}
    G_{\textsc{c}}(x,x')=G_{\textsc{c},0}(x,x')+G_{\textsc{c},\textsc{m}}(x,x')\,.
\end{equation}
The same convention will be applied to other Green's functions we will encounter through the discussions.

The retarded Green's function $G_{\textsc{r}}(x,x')$ is then related to $G_{\textsc{c}}(x,x')$ by
\begin{equation}
    G_{\textsc{r}}(x,x')=\theta(\tau)\,G_{\textsc{c}}(x,x')\,,
\end{equation}
where $\theta(\tau)$ is the Heaviside unit-step function with $\tau=t-t'$. The Hadamard function, defined as the anti-commutator of the field operators, is given by
\begin{equation}
    G_{\textsc{h}}(x,x')=\frac{1}{2}\,\operatorname{Tr}\Bigl(\hat{\rho}_{i}^{(\phi)}\,\bigl\{\hat{\phi}(x),\,\hat{\phi}(x')\bigr\}\Bigr)
\end{equation}
where $\hat{\rho}_{i}^{(\phi)}$ is the density matrix operator of the field in its initial state.

Now consider a neutral atom placed at $\bm{x}$ in the vicinity of the planar boundary, its internal degree of freedom $\chi$ satisfies an operator equation of motion
\begin{equation}\label{E:tbiehd}
    \ddot{\hat{\chi}}(t)+\omega^2_{\textsc{b}}\,\hat{\chi}(t)-\frac{e^2}{m}\int^t_0\!ds\;G_{\textsc{r}}(\bm{x},t;\bm{x},s)\,\hat{\chi}(s)=\frac{e}{m}\,\hat{\phi}_h(\bm{x},t)\,,
\end{equation}
where $e$ is the coupling strength between $\chi$ and the full ambient field $\phi$ obeying the boundary condition. The parameter $m$ is the mass of $\chi$, and $\omega_{\textsc{b}}$ at this moment is identified as the natural frequency before the internal degree of freedom is coupled to the ambient field. For $t>0$, the retarded Green's function $G_{\textsc{r}}(x,x')$ in the nonlocal term in Eq.~\eqref{E:tbiehd} can be replaced by the Pauli-Jordan function $G_{\textsc{c}}(x,x')$. For the massless scalar field, the contribution $G_{\textsc{c},0}(x,x')$ takes the form $-\dfrac{1}{2\pi}\,\partial_t\delta(t-t')$, so the corresponding nonlocal term will give
\begin{align*}
    -\frac{e^2}{m}\int^t_0\!ds\;G_{\textsc{c},0}(\bm{x},t;\bm{x},s)\,\hat{\chi}(s)=\frac{e^2}{2\pi m}\int^t_0\!ds\;\biggl[\frac{\partial}{\partial t}\delta(t-s)\biggr]\,\hat{\chi}(s)=-4\gamma\,\delta(0)\,\hat{\chi}(t)+2\gamma\,\dot{\hat{\chi}}(t)\,.
\end{align*}
where $\gamma=e^2/(8\pi m)$. Thus, it will modify the natural frequency $\omega_{\textsc{b}}$ and introduce a damping term. The frequency correction appears to be divergent. The presence of $\delta(0)$ is a consequence of inclusion of all field modes. In practice or from the perspective of effective field theory, a cutoff frequency $\Lambda$ is introduced to regularize the nonlocal expression in Eq.~\eqref{E:tbiehd}, so $\delta(0)$ is interpreted as $\Lambda/\pi$.

The factor $G_{\textsc{c},\textsc{m}}(x,x')$ on the other hand will account for the nonlocal effect arising from the image of the internat degree of freedom due to the presence of the boundary. In the current configuration, $G_{\textsc{c},\textsc{m}}(x,x')$ takes the form
\begin{equation}\label{E:rbro}
    G_{\textsc{c},\textsc{m}}(x,x')=-G_{\textsc{c},0}(x,\tilde{x}')=-\frac{1}{8\pi z}\,\Bigl[\delta(\tau-2z)-\delta(\tau+2z)\Bigr]\,,
\end{equation}
where $\tilde{\bm{x}}'=(\bm{x}_{\perp},-z)$ since $\bm{x}'=\bm{x}=(\bm{x}_{\perp},z)$. The second term in the square brackets is irrelevant since $\tau>0$. The expression $\delta(\tau-2z)$ encodes the lightlike influence of the atom onto itself due to the planar boundary.

These results allow us to re-cast Eq.~\eqref{E:tbiehd} into the form
\begin{equation}\label{E:iyewed}
    \ddot{\hat{\chi}}(t)+2\gamma\,\dot{\hat{\chi}}(t)+\omega^2_{\textsc{p}}\,\hat{\chi}(t)-\frac{e^2}{m}\int^t_0\!ds\;G_{\textsc{r},\textsc{m}}(\bm{x},t;\bm{x},s)\,\hat{\chi}(s)=\frac{e}{m}\,\hat{\phi}_h(\bm{x},t)\,,
\end{equation}
where we have explicitly kept the contribution due to the presence of the boundary. The physical frequency $\omega_{\textsc{p}}$ is obtained by absorbing into the natural frequency $\omega_{\textsc{b}}$ the correction due to the atom-field interaction.

However, the interpretation of the contribution associated with $G_{\textsc{c},\textsc{m}}(x,x')$ becomes subtle when the atom is very close to the boundary. We see that as $z\to0$, Eq.~\eqref{E:rbro} approaches
\begin{equation}
    G_{\textsc{c},\textsc{m}}(x,x')\to\frac{1}{2\pi}\,\frac{\partial}{\partial t}\delta(\tau)\,,
\end{equation}
which effectively but not entirely counteracts the contribution from $G_{\textsc{c},0}(x,x')$. This implies that in the immediate vicinity of the perfect conducting planar boundary, the internal degree of freedom $\chi$ of the atom will act as a harmonic oscillator that is nearly decoupled from the ambient quantum field. It will oscillate at a frequency slightly shifted from the natural value, and experience a negligible damping, resulting in a relaxation time scale of the internal dynamics significantly longer than if the atom were located further away from the boundary. This phenomenon arises from the Dirichlet boundary condition imposed on the ambient field, which requires the field to vanish at the boundary.

A similar suppression effect between two atoms can be observed when both are placed next to the boundary. The mutual interaction between two atoms will be notably reduced due to the additional interaction between each atom and the image of the other atom.

At late times, the internal dynamics of both atoms, governed by Eqs.~\eqref{E:dkbfo1} and \eqref{E:dkbfo2}, will asymptotically relax to their respective equilibrium states. In this final state, the elements of the covariance matrix attain constant values, which are given explicitly in Eqs.~(A12)--(A14) in the appendix of Ref.~\cite{AFD2} with $\coth\dfrac{\beta\omega}{2}$ replaced by $\operatorname{sgn}(\omega)$ since the ambient field is presumed to be initially in the vacuum state, and with Green's function tensor substituted by those discussed in this Appendix and Ref.~\cite{HHPRD18}. These elements of the covariance matrix then serve as the basis for evaluating the entanglement measures discussed in Sec.~\ref{S:eoeer}.

\section{Entanglement Measure for Gaussian Systems}\label{S:eoeer}

Negativity is a computable, and quantifiable entanglement measure for a {mixed or pure} bipartite Gaussian system. It can be determined from the symplectic eigenvalues of the partially transposed covariance matrix of this bipartite system. Here we include a concise summary of computation and properties relevant to the symplectic eigenvalues of the (partially transposed) covariance matrix.

Let $\bm{R}=(\chi_{1}, p_{1}, \chi_{2}, p_{2})^{T}$, where $(\chi_i,p_i)$ are the canonical variables associated with internal dynamics of atom $i$. The covariance matrix $\bm{\sigma}$ for the bipartite interacting system consisting of two neutral atoms is defined by
\begin{equation}
	\bm{\sigma}=\frac{1}{2}\langle\bigl\{\bm{R},\bm{R}^{T}\bigr\}\rangle=\begin{pmatrix}\bm{A}&\bm{C}\\\bm{C}^{T}&\bm{B}\end{pmatrix}\,,
\end{equation}
where
\begin{align}
	\bm{A}&=\begin{pmatrix}\langle\chi_{1}^{2}\rangle &\dfrac{1}{2}\langle\{\chi_{1},p_{1}\}\rangle\\[10pt]\dfrac{1}{2}\langle\{\chi_{1},p_{1}\}\rangle &\langle p_{1}^{2}\rangle\end{pmatrix}\,,&\bm{B}&=\begin{pmatrix}\langle\chi_{2}^{2}\rangle &\dfrac{1}{2}\langle\{\chi_{2},p_{2}\}\rangle\\[10pt]\dfrac{1}{2}\langle\{\chi_{2},p_{2}\}\rangle &\langle p_{2}^{2}\rangle\end{pmatrix}\,,\\
	\bm{C}&=\begin{pmatrix}\dfrac{1}{2}\langle\{\chi_{1},\chi_{2}\}\rangle &\dfrac{1}{2}\langle\{\chi_{1},p_{2}\}\rangle\\[10pt]\dfrac{1}{2}\langle\{\chi_{2},p_{1}\}\rangle &\dfrac{1}{2}\langle\{p_{1},p_{2}\}\rangle\end{pmatrix}\,.
\end{align}
Here we have assumed $\langle\bm{R}\rangle=0$.

Define
\begin{align}
	\bm{\Omega}&=\begin{pmatrix}\bm{J} &\bm{0}\\\bm{0}&\bm{J}\end{pmatrix}\,,&\bm{J}&=\begin{pmatrix}0 &+1\\-1&0\end{pmatrix}\,,
\end{align}
such that $\bm{\Omega}^{-1}=\bm{\Omega}^{T}=-\bm{\Omega}$, and the canonical commutation relations then read $[\bm{R},\bm{R}^{T}]=i\,\bm{\Omega}$. This allows the Robertson-Schr\"odinger uncertainty principle to be recast in a compact form,
\begin{equation}
	\bm{\sigma}+\frac{i}{2}\,\bm{\Omega}\geq0\,.\label{E:hrvgdjf}
\end{equation}
The $\mathrm{Sp}(2,\mathbb{R})\otimes\mathrm{Sp}(2,\mathbb{R})\subset\mathrm{Sp}(4,\mathbb{R})$ invariants associated with $\bm{\sigma}$ are given by
\begin{align}
	I_{1}&=\det\bm{A}\,,&I_{2}&=\det\bm{B}\,,&I_{3}&=\det\bm{C}\,,&I_{4}&=\operatorname{Tr}\bigl\{\bm{A}\cdot\bm{J}\cdot\bm{C}\cdot\bm{J}\cdot\bm{B}\cdot\bm{J}\cdot\bm{C}^{T}\cdot\bm{J}\bigr\}\,,
\end{align}
with $\det\bm{\sigma}=I_{1}I_{2}+I_{3}^{2}-I_{4}$. In terms of these invariants, the uncertainty principle \eqref{E:hrvgdjf} can be equivalently expressed as the invariant condition,
\begin{equation}
	I_{1}I_{2}+\bigl(\frac{1}{4}-I_{3}\bigr)^{2}-I_{4}\geq\frac{1}{4}\bigl(I_{1}+I_{2}\bigr)\,.
\end{equation}
It is often convenient to introduce another invariant $\Delta(\bm{\sigma})=I_{1}+I_{2}+2I_{3}$. According to Williamson's theorem, there exists a symplectic transformation $\bm{S}$ that brings the covariance matrix $\bm{\sigma}$ into a diagonal form,	
\begin{align}\label{E:ktvkgsf}
	\bm{\sigma}&=\bm{S}^{T}\cdot\bm{K}\cdot\bm{S}\,,&\bm{K}&=\begin{pmatrix}\lambda_{-}\bm{I}_{2} &\bm{0}\\\bm{0} &\lambda_{+}\bm{I}_{2}\end{pmatrix}\,,&\bm{I}_{2}&=\begin{pmatrix}1 &0\\0 &1\end{pmatrix}\,,
\end{align}
where $\lambda_{\pm}$ are the symplectic eigenvalues of $\bm{\sigma}$ with $\lambda_{-}\leq\lambda_{+}$.

By definition of the symplectic transformation $\bm{\Omega}=\bm{S}^{T}\cdot\bm{\Omega}\cdot\bm{S}$, it follows that $\det\bm{S}=1$. Therefore, $\det\bm{\sigma}=\det\bm{K}=\lambda_{-}^{2}\lambda_{+}^{2}$. Moreover, since $\Delta(\bm{\sigma})$ is an $\mathrm{Sp}(2,\mathbb{R})\otimes\mathrm{Sp}(2,\mathbb{R})$ invariant, we have $\Delta(\bm{\sigma})=\lambda_{-}^{2}+\lambda_{+}^{2}$. Combining these results, we obtain
\begin{equation}
	\lambda^{2}_{\pm}=\frac{1}{2}\Bigl[\Delta(\bm{\sigma})\pm\sqrt{\Delta^{2}(\bm{\sigma})-4\det\bm{\sigma}}\Bigr]\,,
\end{equation}
for the covariance matrix $\bm{\sigma}$. According to Eq.~\eqref{E:ktvkgsf}, the uncertainty principle, $\bm{K}+\dfrac{i}{2}\,\bm{\Omega}\geq0$ implies
\begin{align}
	\bigl(\lambda_{+}^{2}-\frac{1}{4}\bigr)\bigl(\lambda_{-}^{2}-\frac{1}{4}\bigr)&\geq0\,,&&\Rightarrow&\lambda_{\pm}&\geq\frac{1}{2}\,.
\end{align}
The symplectic eigenvalues $\lambda_{\pm}$ are positive because $\bm{\sigma}$ is a positive-definite, symmetric real matrix.

Suppose the partial transpose is carried out with respect to atom 2. Partial transpose turns $\bm{R}$ into $\bm{R}'=\bm{\Lambda}\cdot\bm{R}$ with $\bm{\Lambda}=\operatorname{diag}(+1,+1,+1,-1)$. Thus, the covariance matrix after partial transpose, denoted by $\bm{\sigma}^{\textsc{pt}}$, is given by
\begin{equation}
	\bm{\sigma}^{\textsc{pt}}=\bm{\Lambda}\cdot\bm{\sigma}\cdot\bm{\Lambda}=\begin{pmatrix}\bm{A}&\bm{C}'\\\bm{C}'^{T}&\bm{B}\end{pmatrix}\,,
\end{equation}
with $\det\bm{C}'=-\det\bm{C}$. Note that $\bm{\Lambda}$ is not a symplectic matrix, that is, $\bm{\Lambda}\cdot\bm{\Omega}\cdot\bm{\Lambda}\neq\bm{\Omega}$.

For a separable state $\rho$, the partial transpose $\rho^{\textsc{pt}}$ remains a valid positive operator, leading to the uncertainty relation principle,
\begin{equation}
	\bm{\sigma}^{\textsc{pt}}+\frac{i}{2}\,\bm{\Omega}\geq0\,.\label{E:chvbxjhdfs}
\end{equation}	
The aforementioned symplectic invariants are then revised accordingly,	
\begin{equation}
	I_{1}I_{2}+\bigl(\frac{1}{4}-I'_{3}\bigr)^{2}-I'_{4}\geq\frac{1}{4}\bigl(I_{1}+I_{2}\bigr)\,,
\end{equation}	
with $\det\bm{\sigma}^{\textsc{pt}}=I_{1}I_{2}+I'^{2}_{3}-I'_{4}$ and
\begin{align}
	I'_{3}&=\det\bm{C}'\,,&I'_{4}&=\operatorname{Tr}\bigl\{\bm{A}\cdot\bm{J}\cdot\bm{C}'\cdot\bm{J}\cdot\bm{B}\cdot\bm{J}\cdot\bm{C}'^{T}\cdot\bm{J}\bigr\}\,.
\end{align}
It follows from $\bm{\Lambda}^{2}=\bm{I}$ that 
\begin{align*}
    I'_{3}&=-I_{3}\,,&I'_{4}&=I_{4}\,,\\
    \det\bm{\sigma}^{\textsc{pt}}&=\det\bm{\sigma}\,,&\Delta(\bm{\sigma}^{\textsc{pt}})&=I_{1}+I_{2}-2I_{3}=\Delta(\bm{\sigma})-4I_{3}\,.
\end{align*}
Thus, the criterion of separability becomes
\begin{equation}
	I_{1}I_{2}+\bigl(\frac{1}{4}+I_{3}\bigr)^{2}-I_{4}\geq\frac{1}{4}\bigl(I_{1}+I_{2}\bigr)\,,
\end{equation}
which then implies
\begin{equation}
	\lambda^{\textsc{pt}}_{\pm}\geq\frac{1}{2}\,,
\end{equation}
for a separable state, where
\begin{equation}
	(\lambda^{\textsc{pt}}_{\pm})^{2}=\frac{1}{2}\Bigl[\Delta(\bm{\sigma}^{\textsc{pt}})\pm\sqrt{\Delta^{2}(\bm{\sigma}^{\textsc{pt}})-4\det\bm{\sigma}^{\textsc{pt}}}\Bigr]\,.
\end{equation}
Thus, $\lambda^{\textsc{pt}}_{-}<1/2$ signals the existence of entanglement. Hereafter we omit the superscript $\textsc{pt}$ for notational simplicity.

{If two interacting quantum systems are initially prepared in a separable pure state of the combined system, their mutual interaction in general will evolve the combined system into a entangled pure state, symbolically shown as
\begin{align}
    \lvert i_1\rangle\otimes\lvert i_2\rangle\mapsto\sum_k\lvert k_1\rangle\otimes\lvert k_2\rangle
\end{align}
where $\lvert i_1\rangle$, $\lvert i_2\rangle$ are initial states of system 1, 2, and $\lvert k_1\rangle$, $\lvert k_2\rangle$ are possible final states of both systems. If the right hand side cannot be expressed as a single product state, the both systems get entangled. In this case, the state of either one of the systems becomes mixed. The purity $\mu=\operatorname{Tr}\{\rho^2_{\textsc{r}}\}$ and the von Neumann entropy $S_{vN}=-\operatorname{Tr}\{\rho_{\textsc{r}}\ln\rho_{\textsc{r}}\}$, are then measures of the degree of mixedness of the reduced state $\rho_{\textsc{r}}$ of the whole system. For interacting Gaussian systems at late times $t\to\infty$, both purity $\mu$ and the von Neumann entropy $S_{vN}$ can be easily expressed in terms of the late-time determinant, denoted by $\nu^2$, of the covariance matrix of the reduced system, 
\begin{align}
    \mu(\infty)&=\frac{1}{2\nu}\,,&S_{vN}(\infty)&=\Bigl(\nu+\frac{1}{2}\Bigr)\,\ln\Bigl(\nu+\frac{1}{2}\Bigr)-\Bigl(\nu-\frac{1}{2}\Bigr)\,\ln\Bigl(\nu-\frac{1}{2}\Bigr)\,.
\end{align}
Thus, in Paper II and here, we use them to quantify the entanglement between a neutral atom and the ambient quantum field because their calculations are simpler for such a setting. Their properties are briefly used in Sec.~\ref{S:bfbieur}. More details can be found in Ref.~\cite{AFD2}}.


\newpage
\begin{thebibliography}{999}

\bibitem{AFD1}
    J.-T. Hsiang, and B. L. Hu, \textit{Atom-field-medium interactions: Graded influence actions for $N$-harmonic atoms in a dielectric-altered quantum field}, Phys. Rev. A \textbf{110}, 062807 (2024).

\bibitem{AFD2}
    J.-T. Hsiang, and B. L. Hu, \textit{Atom-field-medium interactions. II. Covariance matrix dynamics for $N$ harmonic atoms in a dielectric-altered quantum field and effects of dielectric on atom-field entanglement}, Phys. Rev. A (2025). [arXiv:2503.13022].

\bibitem{QFTop}
    C. J. Isham, \textit{Twisted quantum fields in a curved space-time}, Proc. Roy. Soc. Lond. A \textbf{362}, 383 (1978). 

\bibitem{QFTop1}
    S. J. Avis, and C. J. Isham, \textit{Vacuum solutions for a twisted scalar field}, Proc. Roy. Soc. Lond. A \textbf{363}, 581 (1978).

\bibitem{QFTop2}
    J. S. Dowker, and R. Banach, \textit{Quantum field theory on Clifford-Klein space-times. The
Effective Lagrangian and vacuum stress energy tensor}, J. Phys. A \textbf{11}, 2255 (1978). 

\bibitem{QFTop3}
    B. S. DeWitt, C. F. Hart, and C. J. Isham, \textit{Topology and quantum field theory}, Physica A \textbf{96}, 197 (1979).

\bibitem{QFTop4}
    For an example of how topology enters in the quantum field-mediated two-atom entanglement dynamics, see, S.-Y. Lin, C. H. Chou, and B. L. Hu,  \textit{Entanglement dynamics of detectors in an Einstein cylinder},  JHEP \textbf{03}, 047 (2016).

\bibitem{Casimir} 
    H. B. G. Casimir, \textit{On the attraction between two perfectly conducting plates}, Kon. Ned. Akad. Wetensch. Proc. \textbf{51}, 793 (1948). 

\bibitem{Casimir1}
    M. Bordag, U. Mohideen, and V. M. Mostepanenko, \textit{New developments in the Casimir effect}, Phys. Rep. \textbf{353}, 1 (2001).

\bibitem{Casimir2}
    G. L. Klimchitskaya, U. Mohideen, and V. M. Mostepanenko, \textit{The Casimir force between real materials: Experiment and theory}, Rev. Mod. Phys. \textbf{81}, 1827 (2009).

\bibitem{Casimir3}
    D. Dalvit, P. Milonni, D. Roberts, and F. Da Rosa, \textsl{Casimir Physics}, Lecture Notes in Physics, Vol. 834 (Springer-Verlag, Heidelberg, 2011).

\bibitem{Casimir4}
    K. A. Milton, \textsl{State of The Quantum Vacuum, The: Casimir Physics in the 2020's} (World Scientific, Singapore, 2022).

\bibitem{QFCS}
    N. D. Birrell, and P. C. W. Davies, {\sl Quantum Fields in Curved Spaces} (Cambridge University Press, Cambridge, 1982).

\bibitem{QFCS1}
    L. Parker, and D. Toms, \textsl{Quantum Field Theory in Curved Spacetime: Quantized Fields and Gravity} (Cambridge University Press, Cambridge, 2009).

\bibitem{QFCS2}
    B. L. Hu, and E. Verdaguer, {\sl Semiclassical and Stochastic Gravity -- Quantum Field Effects on Curved Spacetime} (Cambridge University Press, Cambridge, 2020).

\bibitem{DCE}  
    V. V. Dodonov, \textit{Dynamical Casimir effect: 55 years later}. Physics \textbf{7}, 10 (2025).

\bibitem{CPC} 
    Many important references are cited  in  this recent overview:  J.-T. Hsiang, and B. L. Hu, \textit{Foundational issues in dynamical Casimir effect and analogue features in cosmological particle creation}, Universe \textbf{10}, 418 (2024).

\bibitem{Unr76} 
    W. G. Unruh, \textit{Notes on black-hole evaporation}, Phys. Rev. D \textbf{14}, 870 (1976).

\bibitem{Unr761} 
    L. C. B. Crispino, A. Higuchi, and G. E. A. Matsas, \textit{The Unruh effect and its applications}, Rev. Mod. Phys. \textbf{80}, 787 (2008).

\bibitem{Haw74}  
    S. W. Hawking, \textit{Black hole explosions?}, Nature \textbf{248}, 30 (1974).

\bibitem{Schroedinger} 
    E. Schr\"odinger, \textit{The present status of quantum mechanics}, Die Naturwissenschaften, \textbf{23}, 1 (1935).

\bibitem{RQI} See isrqi.net

\bibitem{Peres}  
    A. Peres, and D. R. Terno, \textit{Quantum information and relativity theory}, Rev. Mod. Phys. \textbf{76}, 93 (2004).

\bibitem{EntQG} 
    See, e.g., M. Van Raamsdonk, \textit{Lectures on gravity and entanglement} in \textsl{New Frontiers in Fields and Strings: TASI 2015 Proceedings of the 2015 Theoretical Advanced Study Institute in Elementary Particle Physics}. 

\bibitem{AnalogG} 
    C. Barcelo, S. Liberati, and M. Visser, \textit{Analogue gravity}, Liv. Rev. Rel. \textbf{14}, 1 (2011).

\bibitem{DeW79} 
    B. S. DeWitt, in \textit{General Relativity: an Einstein Centenary Survey}, edited by S. W. Hawking and W. Israel (Cambridge University Press, Cambridge, 1979).

\bibitem{MovMir} 
    G. T. Moore, \textit{Quantum theory of the electromagnetic field in a variable-length one-dimensional cavity}, J. Math. Phys. \textbf{11}, 2679 (1970).

\bibitem{MovMir1} 
    P. C. W. Davies, and S.A. Fulling, \textit{Radiation from moving mirrors and from black holes}, Proc. Roy. Soc. Lond. A \textbf{356}, 237 (1977). 

\bibitem{MovMir2}
    S. A. Fulling, and P. C. W. Davies, \textit{Radiation from a moving mirror in two dimensional space-time: conformal anomaly}, Proc. R. Soc. Lond. A \textbf{348}, 393 (1976).

 \bibitem{MovMir3}
    W. R.Walker, \textit{Particle and energy creation by moving mirrors}, Phys. Rev. D \textbf{31}, 767 (1985).

\bibitem{MovMir4}
    R. D. Carlitz, and R. S. Willey, \textit{Reflections on moving mirrors}, Phys. Rev. D \textbf{36}, 2327 (1987).

\bibitem{2AEntT} 
    There is an abundance of references on atom and ion entanglement. For early experimental work, see, e.g., S. B. Zheng, and G. C. Guo, \textit{Efficient scheme for two-atom entanglement and quantum information processing in cavity QED}, Phys. Rev. Lett. \textbf{85}, 2392 (2000).  
    
\bibitem{2AEntT1} 
    R. Blatt, and D. Wineland, \textit{Entangled states of trapped atomic ions}. Nature, \textbf{453}, 1008 (2008).

\bibitem{2AEntT2}
    An early theoretical review is Z. Ficek, and R. Tana\'s, \textit{Entangled states and collective nonclassical effects in two-atom systems}, Phys. Rep. \textbf{372}, 369 (2002). 
    
\bibitem{2AEntT3}
    For early theoretical work which pointed out the importance of nonMarkovian effects, see   C. Anastopoulos,  S. Shresta and B. L. Hu, \textit{Quantum entanglement under non-Markovian dynamics of two qubits interacting with a common electromagnetic field},  [arXiv:quant-ph/0610007].  
    
\bibitem{2AEntT4}
    B. Bellomo, R. Lo Franco, and G. Compagno, \textit{Non-Markovian effects on the dynamics of entanglement}, Phys. Rev. Lett. \textbf{99}, 160502 (2007).

\bibitem{2AEntE}
    D. Braun, \textit{Creation of entanglement by interaction with a common heat bath}, Phys. Rev. Lett. \textbf{89}, 277901 (2002).

\bibitem{2AEntE1}
    H. T. Dung, S. Scheel, D.-G. Welsch, and L. Kn\"oll, \textit{Atomic entanglement near a realistic microsphere}, J. Opt. B \textbf{4}, S169 (2002).
    
\bibitem{2AEntE2}
    F. Benatti, R. Floreanini, and M. Piani, \textit{Environment induced entanglement in markovian dissipative dynamics}, Phys. Rev. Lett. \textbf{91}, 070402 (2003).

\bibitem{CasPol} 
     H. B. G. Casimir, and D. Polder, \textit{The influence of retardation on the London-van der Waals forces}, Phys. Rev. \textbf{73}, 360 (1948).

\bibitem{Kanu24} 
    M. Izadyari, O. Pusuluk, K. Sinha,  and \"O. E. M\"ustecapl{\i}o{\u g}lu,  \textit{Steady-state entanglement generation via Casimir-Polder interactions},  [arXiv:2406.02270v3].

\bibitem{Rong} 
    R. Zhou, R. O. Behunin, S.-Y. Lin, and B. L. Hu, \textit{Boundary effects on quantum entanglement and its dynamics in a detector-field system}, JHEP \textbf{08}, 040 (2013).

\bibitem{Ent2UD} 
    J. Hu, and H. Yu, \textit{Entanglement dynamics for uniformly accelerated two-level atoms},  Phys. Rev. A \textbf{91}, 012327 (2015). 

\bibitem{Ent2UD1} 
    W. Zhou, R. Passante, and L. Rizzuto,  \textit{Resonance interaction energy between two accelerated identical atoms in a coaccelerated frame and the Unruh effect},  Phys. Rev. D \textbf{94}, 105025 (2016).

\bibitem{Ent2UD2} 
    M. S. Soares, G. Menezes and  N. F. Svaiter, \textit{Entanglement dynamics: Generalized master equation for uniformly accelerated two-level systems},  Phys. Rev. A \textbf{106}, 062440 (2022). 

\bibitem{Ent2UD3} 
    M. S. Soares, G. Menezes, and  N. F. Svaiter,  \textit{A dynamic theory of entanglement for uniformly accelerated atoms}  Class. Quantum Grav.  \textbf{42}, 115013 (2025).

\bibitem{KapTjo23} 
    G. Kaplanek, and E. Tjoa,  \textit{Effective master equations for two accelerated qubits},  Phys. Rev. A \textbf{107}, 012208 (2023).

\bibitem{CalLeg83} 
    A. O. Caldeira, and A. J. Leggett, \textit{Path integral approach to quantum Brownian motion}, Physica A \textbf{121}, 587 (1983).

\bibitem{HomaCL}  
    G. Homa, J. Z. Bernad, and L. Lisztes, \textit{Positivity violations of the density operator in the Caldeira-Leggett master equation}, Eur. Phys. J. D \textbf{73}, 53 (2019); Erratum: Eur. Phys. J. D \textbf{73}, 128 (2019).

\bibitem{Lindblad}  
    G. Lindblad, \textit{On the generators of quantum dynamical semigroups}, Commun. Math. Phys. \textbf{48}, 119 (1976).

\bibitem{GKS} 
    V. Gorini, A. Kossakowski, and E. C. G. Sudarshan, \textit{Completely positive dynamical semigroups of $N$-level systems}, J. Math. Phys. \textbf{17}, 821 (1976).

\bibitem{LinbPatho} 
    F. Haake, and R. Reibold, \textit{Strong damping and low-temperature anomalies for the harmonic oscillator}, Phys. Rev. A \textbf{32}, 2462 (1985).

\bibitem{LinbPatho1} 
    B. M. Garraway, \textit{Nonperturbative decay of an atomic system in a cavity}, Phys. Rev. A \textbf{55}, 2290 (1997).

\bibitem{LinbPatho2} 
    A. Lampo, S. H. Lim, J. Wehr, P. Massignan, and M. Lewenstein, \textit{Lindblad model of quantum Brownian motion}, Phys. Rev. A \textbf{94}, 042123 (2016).

\bibitem{LinHu06}  
    S.-Y. Lin, and B. L. Hu, \textit{Accelerated detector--quantum field correlations: From vacuum fluctuations to radiation flux}, Phys. Rev. D \textbf{73} (2006).
 
\bibitem{LinHu07}      
    S.-Y. Lin, and B. L. Hu, \textit{Backreaction and Unruh effect: New insights from exact solutions of uniformly accelerated detectors}, Phys. Rev. D \textbf{76} (2007).

\bibitem{LinHu09}   
    S.-Y. Lin, and B. L. Hu, \textit{Temporal and spatial dependence of quantum entanglement from field theory perspective}, Phys. Rev. D \textbf{79}, 085020 (2009).

\bibitem{MOF2}  
    K. Sinha, S.-Y. Lin, and B. L. Hu, \textit{Mirror-field entanglement in a microscopic model for quantum optomechanics}, Phys. Rev. A \textbf{92}, 023852 (2015).

\bibitem{HHPRD16}   
    J.-T. Hsiang, and B. L. Hu, \textit{Distance and coupling dependence of entanglement in the presence of a quantum field}, Phys. Rev. D \textbf{92}, 125026 (2015).

\bibitem{HHPRD18}
    J.-T. Hsiang, Y. Suba\c{s}{\i}, C. H. Chou, and B. L. Hu, \textit{Quantum thermodynamics from the nonequilibrium dynamics of open systems: energy, heat capacity and the third law}, Phys. Rev. E \textbf{97}, 012135 (2018).

\bibitem{HH19}
    J.-T. Hsiang, and B. L. Hu, \textit{Atom-field interaction - From vacuum fluctuations to quantum radiation and quantum dissipation or radiation reaction}, Physics \textbf{1}, 430 (2019).

\bibitem{BHH23}
    M. Bravo, J.-T. Hsiang, and B. L. Hu, \textit{Fluctuations-induced quantum radiation and reaction from an atom in a squeezed quantum field}, Physics \textbf{5}, 554 (2023).

\bibitem{Rohrlich}
	F. Rohrlich, \href{https://doi.org/10.1142/6220}{\textsl{Classical Charged Particles, 3rd Edition}} (World Scientific, Singapore, 2007).

\bibitem{GHW09}
	S. E. Gralla, A. I. Harte, and R. M. Wald, \href{https://doi.org/10.1103/PhysRevD.80.024031}{\textit{Rigorous derivation of electromagnetic self-force}}, Phys. Rev. D \textbf{80}, 024031 (2009).
	
\bibitem{BDR19}
	C. Bild, D.-A. Deckert, and H. Ruhl, \href{https://doi.org/10.1103/PhysRevD.99.096001}{\textit{Radiation reaction in classical electrodynamics}}, Phys. Rev. D \textbf{99}, 096001 (2019).
	
\bibitem{Yaghjian}	
	A. Yaghjian, \href{https://doi.org/10.1007/b98846}{\textsl{Relativistic Dynamics of a Charged Sphere - Updating the Lorentz-Abraham Model}} (Springer-Verlag, New York, 2006).

\bibitem{Erber61}
	T. Erber, \href{https://doi.org/10.1002/prop.19610090702}{\textit{The classical theories of radiation reaction}}, Fortschr. Phys. \textbf{9}, 343 (1961).

\bibitem{MS74}
	E. H. Moniz, and D. H. Sharp, \href{https://doi.org/10.1103/PhysRevD.10.1133}{\textit{Absence of runaways and divergent self-mass in nonrelativistic quantum electrodynamics}}, Phys. Rev. D \textbf{10}, 1133(1974).

\bibitem{FOC91} 
	G. W. Ford, and R. F. O'Connell, \href{https://doi.org/10.1016/0375-9601(91)90054-C}{\textit{Radiation reaction in electrodynamics and the elimination of runaway solutions}}, Phys. Lett. A \textbf{157}, 217 (1991).

\bibitem{PJM82}
	L. de la Pe\~na, J. L. Jim\'enez, and R. Montemayor, \href{https://doi.org/10.1007/BF02721242}{\textit{The classical motion of an extended charged particle revisited}}, Nuovo Cim. B \textbf{69}, 71 (1982).

\bibitem{GLR10}
	C. R. Galley, A. K. Leibovich, and I. Z. Rothstein, \href{https://doi.org/10.1103/PhysRevLett.105.094802}{\textit{Finite size corrections to the radiation reaction force in classical electrodynamics}}, Phys. Rev. Lett. \textbf{105}, 094802 (2010).

\bibitem{LS19} 
	A. Lanir, and O. Sela, \href{https://doi.org/10.1103/PhysRevD.99.064031}{\textit{Curing the self-force runaway problem in finite-difference integration}}, Phys. Rev. D \textbf{99}, 064031 (2019).

\bibitem{HL08}
    J.-T. Hsiang, and T.-H. Wu, and D.-S. Lee, \href{https://doi.org/10.1103/PhysRevD.77.105021}{\textit{Stochastic Lorentz forces on a point charge moving near the conducting plate}}, Phys. Rev. D \textbf{77}, 105021 (2008).

\bibitem{HH22}
    J.-T. Hsiang, and B. L. Hu, \textit{Non-Markovian Abraham-Lorentz-Dirac equation: Radiation reaction without pathology}, Phys. Rev. D \textbf{106}, 125018 (2022).

\bibitem{HAH22}
    J.-T. Hsiang, O. Ar{\i}soy, and B. L. Hu, \textit{Entanglement dynamics of coupled quantum oscillators in independent nonMarkovian baths}, Entropy \textbf{24}, 1814 (2022).

\bibitem{AHH23}
    O. Ar{\i}soy, J.-T, Hsiang, and B.-L. Hu, \textit{Hot entanglement? - Parametrically coupled quantum oscillators in two heat baths: instability, squeezing and driving}, JHEP \textbf{2023}, 122 (2023).

\bibitem{DodHal}
    P. J. Dodd, and J. J. Halliwell, \textit{Disentanglement and decoherence by open system dynamics}, Phys. Rev. A \textbf{69}, 052105 (2004).
    
\bibitem{YuEberly1}
    T. Yu, and J. H. Eberly, \textit{Finite-time disentanglement via spontaneous emission}, Phys. Rev. Lett. \textbf{93}, 140404 (2004).

\bibitem{Almeida}
    M. P. Almeida,  F. de Melo, M. Hor-Meyll, A. Salles, S. P. Walborn, P. H. Souto Ribeiro, and L. Davidovich, \textit{Environment-induced sudden death of entanglement}, Science \textbf{316}, 579 (2007).
    
\bibitem{YuEberly2}
    T. Yu, and J. H. Eberly, \textit{Sudden death of entanglement}, Science \textbf{323}, 598 (2009).
\end{thebibliography}
\end{document}